\documentclass[letterpaper,11pt]{article}
\pdfoutput=1

\usepackage{jheppub}
\usepackage{multirow}
\usepackage{braket}
\usepackage{mathrsfs}
\usepackage{tikz}
\usepackage{subfig}
\usepackage{xspace}
\usepackage{xcolor}
\usepackage{float,color}
\usepackage{siunitx}
\usepackage[countmax]{subfloat}

\providecommand{\href}[2]{#2}

\definecolor{darkred}{rgb}{0.5,0.0,0.0}
\definecolor{darkblue}{rgb}{0.0,0.0,0.9}
\definecolor{darkerblue}{rgb}{0.0,0.0,0.5}
\definecolor{darkgreen}{rgb}{0.0,0.5,0.0}
\definecolor{black}{rgb}{0.0,0.0,0.0}
\definecolor{brown}{rgb}{0.6,0.4,0.2}

\title{\boldmath Deep learning in color: towards automated quark/gluon jet discrimination}

\preprint{ 
\begin{flushright}
MIT--CTP 4866
 \end{flushright}}

\author[a]{Patrick T. Komiske,}
\author[a]{Eric M. Metodiev,}
\author[b]{and Matthew D. Schwartz}

\affiliation[a]{Center for Theoretical Physics, Massachusetts Institute of Technology, Cambridge, MA 02139, USA}
\affiliation[b]{Department of Physics, Harvard University, Cambridge, MA 02138, USA}

\emailAdd{pkomiske@mit.edu}
\emailAdd{metodiev@mit.edu}
\emailAdd{schwartz@physics.harvard.edu}

\abstract{Artificial intelligence offers the potential to automate challenging data-processing tasks in collider physics. To establish its prospects, we explore to what extent deep learning with convolutional neural networks can discriminate quark and gluon jets better than observables designed by physicists. Our approach builds upon the paradigm that a jet can be treated as an image, with intensity given by the local calorimeter deposits. We supplement this construction by adding color to the images, with red, green and blue intensities given by the transverse momentum in charged particles, transverse momentum in neutral particles, and pixel-level charged particle counts. Overall, the deep networks match or outperform traditional jet variables. We also find that, while various simulations produce different quark and gluon jets, the neural networks are surprisingly insensitive to these differences, similar to traditional observables. This suggests that the networks can extract robust physical information from imperfect simulations.}

\begin{document} 
\maketitle
\flushbottom

\section{Introduction}
High energy particle collisions produce an enormous amount of data. For example, the Large Hadron Collider (LHC) is currently generating petabytes per year. Sorting through all of this data is a herculean task, but one that should be amenable to processing using modern developments in data science and artificial intelligence. Neural networks and other approaches already play a significant role in LHC data processing, particularly at the lowest levels, in the electronics~\cite{Aad:2014yva}, or in intermediate level tasks such as tagging of bottom quarks~\cite{Aad:2015ydr,Tosi2015} or tau identification~\cite{CMS2012a}, and in matching data to non-perturbative physics, such as in fitting parton distribution functions~\cite{Ball:2014uwa}. They have also been used effectively for distinguishing certain signals from their known backgrounds~\cite{Maggipinto:1997eh, Gallicchio:2010dq, CMS2013, Baldi:2014pta}. For these applications, one generally constructs a set of physically-motivated but often highly-correlated observables, such as the dijet mass, or angular distributions of decay products, and the neural network is used to combine them into a single discriminant. One might imagine, however, that such an approach is limited by the creativity of physicists who construct the input observables to begin with. Thus, it is important to determine how well neural networks can do at discriminating two event samples with minimal physical input. In particular, in this paper we explore whether artificial intelligence can do well at the challenging task of distinguishing quark jets from gluon jets using data in reasonably raw form rather than using carefully constructed observables.

An arguably minimal approach to processing the LHC data is the ``jet images'' approach introduced in~\cite{Cogan:2014oua,Almeida:2015jua}. The idea behind jet images is to treat the energy deposits in a calorimeter as intensities in a 2D image. Then one can apply sophisticated algorithms developed for image recognition to particle physics. This and related neural network approaches were used for boosted $W$ boson tagging in~\cite{Cogan:2014oua} and~\cite{deOliveira:2015xxd}, top tagging in~\cite{Almeida:2015jua}, heavy-flavor tagging in~\cite{Baldi:2016fql,Guest:2016iqz}, and comparing parton shower uncertainties in~\cite{Barnard:2016}. In many of these studies the data was manipulated using some physical insight before being sent into the network. For example, boosted hadronically-decaying $W$ bosons generally look like large ``fat jets" with two fairly well-defined subjets. Using this insight, in~\cite{Cogan:2014oua},  each jet image was rotated to align with a moment of the fat jet. While some pre-processing is always useful to make the network training more efficient, we will attempt to avoid any pre-processing motivated by physical insight into the samples. For example, we allow generic pre-processing, like normalizing the pixel intensities~\cite{kriz2012}, but avoid steps like looking for subjets that we expect in signal samples but not in background samples.

One of the most challenging tasks in collider physics is to tell apart jets initiated by light quarks from those initiated by gluons. This problem has been studied for decades~\cite{Pumplin:1991kc,Lonnblad:1990,Acton:1993, Alexander:1991} with a fair amount of recent activity~\cite{deLime:2016,Larkoski:2014,Bhattacherjee:2015}. Gluon acceptance efficiencies from 20\% to 5\% are achievable at a 50\% quark acceptance working point~\cite{Gallicchio:2012ez}. The large variation is a result of assumptions about detector properties  (better angular resolution produces better discrimination) and which simulations are used to approximate the data (e.g. quark and gluon jets are more distinguishable in {Pythia} than in {Herwig}).  Some general lessons from~\cite{Gallicchio:2012ez} and~\cite{Gallicchio:2011xq} were that there are essentially two complementary types of observables: shape and count observables. Shape observables are quantities such as the width or girth of a jet, its mass, or an energy-correlation function; count observables are quantities such as the number of charged particles in a jet or the number of distinct calorimeter cells that are triggered. A recent study~\cite{Badger:2016bpw} explored the discrepancy among the simulations, finding that programs with more sophisticated parton showers, such as {Vincia}, tend to perform intermediately between {Herwig} and {Pythia}.  In studies with actual data~\cite{Aad:2014gea}, the relevant observables also seemed to fall somewhere in between {Herwig} and {Pythia}, suggesting that the improved parton shower models may produce more accurate simulations. 

From these studies, one may draw a couple of critical observations. First, current simulations of quark and gluon jet properties are not completely trustworthy. This naturally suggests that one should use discriminants with a solid theoretical justification, so that one does not have to rely on the simulations. The set of observables constructed in~\cite{Gallicchio:2011xq,Gallicchio:2012ez}, such as girth and track count, were all motivated physically and so one expects them to work on data whether or not the simulations agree with the data.  A typical semi-classical theory argument is that gluon jets should have about twice as much radiation as quark jets ($C_A/C_F = 9/4$), making them wider and with more particles.  This Casimir scaling only goes so far, however, as it does not tell us if track count and girth should be complementary or not. Moreover, detector effects and hadronization have an important effect on the jet substructure that is difficult to approach analytically. 

A second observation is that the theory and simulations seem to be improving. For example, it is already possible to make trustworthy calculations beyond the semi-classical limit (e.g. in~\cite{Frye:2016aiz} it was shown that soft-drop allows for an unambiguous quark/gluon jet definition). First-principles calculations of correlations among observables are also being explored~\cite{Larkoski:2014}. Thus, it is easy to imagine that the simulations will be trustworthy before long. Therefore, our approach in this paper will be to pretend that we live in the future, where the simulations are in fantastic agreement with data. In such an ideal world, are physically motivated observables still necessary, or can artificial intelligence, through deep neural networks, truly find an optimal solution to the quark/gluon discrimination problem?

This paper is intended to be readable by an audience with minimal previous exposure to deep learning. We begin in Section~\ref{sec:deep} with an introduction to some of the terminology used in the neural network community and an overview of some of the insights from recent years that have led to the rapid development of this technology. In Section~\ref{sec:events} we discuss our data simulation and network architecture, including our innovation of adding multiple channels (``colors'') to the jet images.  Section~\ref{sec:fisher} explores the fisher-jet approach, following~\cite{Cogan:2014oua} and its connection to convolutional filters. The network performance is discussed in Section~\ref{sec:performance} and our conclusions are presented in Section~\ref{sec:conc}.

\section{Deep Neural Networks \label{sec:deep}}
Artificial neural networks (ANNs or  NNs) are a powerful tool in machine learning and have been successfully applied to many problems in fields such as computer vision~\cite{kriz2012}, natural language processing~\cite{coll2008}, and physics~\cite{bald2014, diel2015}. Recent comprehensive introductions to neural networks and deep learning can be found in~\cite{niel2015}~and~\cite{good2016}.

The  basic goal of a neural network is to learn a function from a set of fixed-size inputs to a fixed-size output. The network consists of the {\it input layer}, one or more {\it intermediate} or {\it hidden} layers consisting of a set of {\it units} or {\it nodes}, and an {\it output layer}. In a {\it feed-forward} neural network, the layers are ordered and each unit of a layer connects to some subset of the units of previous layers. A layer is {\it dense} if each of its units connects to all of the units in the previous layer.  A network is {\it deep} if there are several hidden layers. Each connection between units in adjacent layers has a {\it weight} and each unit has a {\it bias}. 

The value of a unit in a non-input layer is obtained by multiplying the values of its inputs by the respective weights of their connections, summing these values, and adding the bias. This sum is then operated on by an {\it activation function}. The idea of an activation function is inspired by  biological neurons that  fire after a certain threshold is reached. Correspondingly, these functions were traditionally taken to be smoothed-out step functions, like a sigmoid or  logistic function. One of the insights which allowed the rapid progress in deep learning in recent years is the observation that training can be easier with different activation functions.

For example, the sigmoid has a nearly-vanishing gradient for inputs far from zero, which can lead to {\it saturation}, whereby the network becomes insensitive to changes in input unit values. To avoid this, modern applications in computer vision typically use the rectified linear unit (ReLU)~\cite{relu}, with $\text{ReLU}(x)=\max\{0,x\}$, for an activation function. ReLUs are computationally fast to evaluate and avoid saturation since their derivative is 1 for any positive value of the input.  

To learn a function with a neural network, the weights and biases  are typically determined through {\it supervised learning}, whereby the network is shown many examples of the input for which the true value of the function is known. In the case of classification, the network is shown examples for which the true class is known. A {\it loss function} encapsulates the difference between the network output and the true class. Minimization of the loss function proceeds by calculating the gradient of the loss function with respect to the weights and biases of the network using the {\it back-propagation algorithm}, and updating these parameters via stochastic gradient descent. In this way, the network is trained to classify new examples.

 It can be proven that even a network with as few as one hidden layer (a shallow network) can approximate any well-behaved function to arbitrary accuracy if it has sufficiently many units~\cite{horn1989}. Deep networks offer the potential to approach this optimum much more efficiently, by having more layers with fewer units, than dense, shallow networks~\cite{schm2015, szeg2015}. The additional hidden layers allow the network to identify low level features in the early layers and more abstract, higher-level features in the later  layers.

 Until recently, complications such as long training times for large datasets, unit saturation, and overfitting have prevented the effective usage of deep networks. The increasing availability of computing power, especially specialized graphics processing units (GPUs), has sped up training times of deep networks. The introduction of new activation functions such as the ReLU have alleviated saturation issues by avoiding the vanishing gradient problem. Another problem with in machine learning is {\it overfitting}, where a model picks up on overly-fine details of the training samples and performs poorly on test samples. The term {\it regularization} refers to methods introduced to avoid overfitting. An effective regularization method for neural networks is {\it dropout}~\cite{dropout}, in which a random subset of units are ignored during each training update in order to avoid over-dependence on any particular part of the network. Dropout works well to regularize large and complex networks.

Another important development has been the adoption of convolutional neural networks (CNNs), namely networks which include one or more \emph{convolutional layers}, as a standard tool in image recognition problems. A convolutional layer is computed from the previous layer by convolving with a filter. A {\it filter} is an $n\times n$ grid of weights. The convolution multiplies this filter by a patch of the previous layer, sums the values, and adds a bias, and then applies the activation function. The filter is shifted along the image by a {\it stride length}, usually taken to be 1 unit. For instance starting with a $20\times 20$ pixel image, a $5\times 5$ filter can be placed at $16\times 16$ locations on the image, so with a stride length of 1 the convolutional layer has $16\times 16$ units. Note that the same filter is used for each offset, so only the $5\times 5$ weights in the filter and the bias are independently trained. Typically many different filters connected to the same inputs are trained, starting from different initial values (in our application, each convolutional layer will have 64 filters). Different filters can pick up on different and often complementary aspects of the image. 

Finally, CNNs typically have  {\it max-pooling} layers. These layers are simply down-samplings, where each unit takes the the maximum value in $m\times m$ patches in a convolutional layer. Maxpooling reduces the number of parameters, which effectively allows the network to focus on relevant features. For instance, a $16\times16$ unit convolutional layer would be  downsampled by  $2\times2$ maxpooling to an $8\times8$ unit layer. This reduced layer is then  taken as input into the next layer of the network.

\section{Event Generation and Network Architecture \label{sec:events}}
Previous studies of quark and gluon jet discrimination have found a notable difference among different simulations~\cite{Gallicchio:2012ez,Badger:2016bpw}. Thus we generated events using both {Pythia} 8.219~\cite{Pythia8.2:2015} and {Herwig} 7.0~\cite{Herwig:2008, Herwig:2016} using their respective default tunings and shower parameters. We simulated dijet events in $pp$ collisions at  $\sqrt{s}=\SI{13}{TeV}$. The light-quark ($u,d,s$) initiated jets come from parton-level hard processes $qq \to qq$, $q\bar{q} \to q\bar{q}$ or $gg\to qq$; the gluon-initiated jets come from $gg\to gg$ or $q\bar q\to gg$ for gluon production. We turned off $qg \to q g$ for simplicity.  Final state particles with pseudorapidity $|\eta|< 2.5$ were kept and neutrinos were discarded. The resulting events were then clustered with {FastJet}~\cite{FastJet:2012} using the anti-$k_T$ algorithm~\cite{cacc2008} with a jet radius of $R=0.4$

Four jet-transverse-momentum ($p_T$) ranges were used: $100-$\SI{110}{GeV}, $200-$\SI{220}{GeV}, $500-$\SI{550}{GeV}, $1000-$\SI{1100}{GeV}. The parton-level $p_T$ cuts were chosen 20\% broader than these ranges in order to ensure that jet $p_T$ distributions were not distorted from distributions with no parton-level cuts. For each $p_T$ range, 100k each of quark/gluon Pythia jets were generated. For the $200-$\SI{220}{GeV} and $1000-$\SI{1100}{GeV} ranges, 100k each of quark/gluon Herwig jets were generated for a comparison of different MCs. The four $p_T$ ranges will be referred to by their lower limits: \SI{100}{GeV}, \SI{200}{GeV}, \SI{500}{GeV}, \SI{1000}{GeV}. In each set of 100k events, 90k were used for training and 10k for testing. 

For each jet in a given $p_T$ range, we constructed a jet image, following~\cite{Cogan:2014oua, deOliveira:2015xxd}. The images are square arrays in $(\eta,\phi)$-space with each pixel given by the total $p_T$  deposited in the associated region. The images are used as the input layer to the neural network. It is helpful to have a center pixel, so we chose odd numbers of grid lengths, e.g. $33\times 33$ pixels. For a jet radius $R$, the image has size $2R\times2R$. In this paper we used $R=0.4$, which for $33\times33$ pixels corresponds to a $\Delta \eta = \Delta\phi = 0.024$. The discretization into $\Delta\eta\times\Delta\phi = 0.024\times 0.024$ pixels also acts as a kind of coarse-graining often used as a primitive detector simuilator (e.g. see~\cite{Kaplan:2008ie}). We follow~\cite{deOliveira:2015xxd} in using transverse momentum as our pixel intensity rather than the energy due  to the invariance of the transverse momentum under longitudinal boosts.

The approach taken in this paper regards the input values as exact, neglecting issues related to measurement uncertainty. Unlike in typical computer vision applications, the inputs to neural networks in physics applications have uncertainties associated with them. Neural networks are capable of sensible error propagation when supplied with input values with errors. One approach taken by the NNPDF collaboration~\cite{NNPDF:2009} is to sample several ``replicated'' datasets from the distribution determined by the measured data with errors and to train a separate network on each of them. Then the uncertainty of a prediction on a new data point can be determined from the distribution of model outputs on that data point.

\subsection{Pre-processing}
The following series of data-motivated pre-processing steps were applied to the jet images:
\begin{enumerate}
\item {\bf Center:} Center the jet image by translating in ($\eta,\phi$) so that the total $p_T$-weighted centroid pixel is at $(\eta,\phi) = (0,0)$. This operation corresponds to rotating and boosting along the beam direction to center the jet.
\item {\bf Crop:} Crop to a $33\times 33$ pixel region centered at $(\eta,\phi) = (0,0)$, which captures the region with $\eta,\phi \in (-R,R)$ for $R = 0.4$.
\item {\bf Normalize:} Scale the pixel intensities such that $\sum_{ij} I_{ij} = 1$ in the image, where $i$ and $j$ index over the pixels.
\item {\bf Zero-center:}\label{step:zc} Subtract the mean $\mu_{ij}$ of the normalized training images from each image, transforming each pixel intensity as $I_{ij} \to I_{ij} - \mu_{ij}$.
\item {\bf Standardize:} \label{step:stand} Divide each pixel value by the standard deviation $\sigma_{ij}$ of that pixel value in the normalized training dataset, $I_{ij} \to I_{ij}/(\sigma_{ij} + r)$. A value of $r = 10^{-5}$ was used to suppress noise.
\end{enumerate}

\begin{figure}
\centering
\includegraphics[scale=0.7]{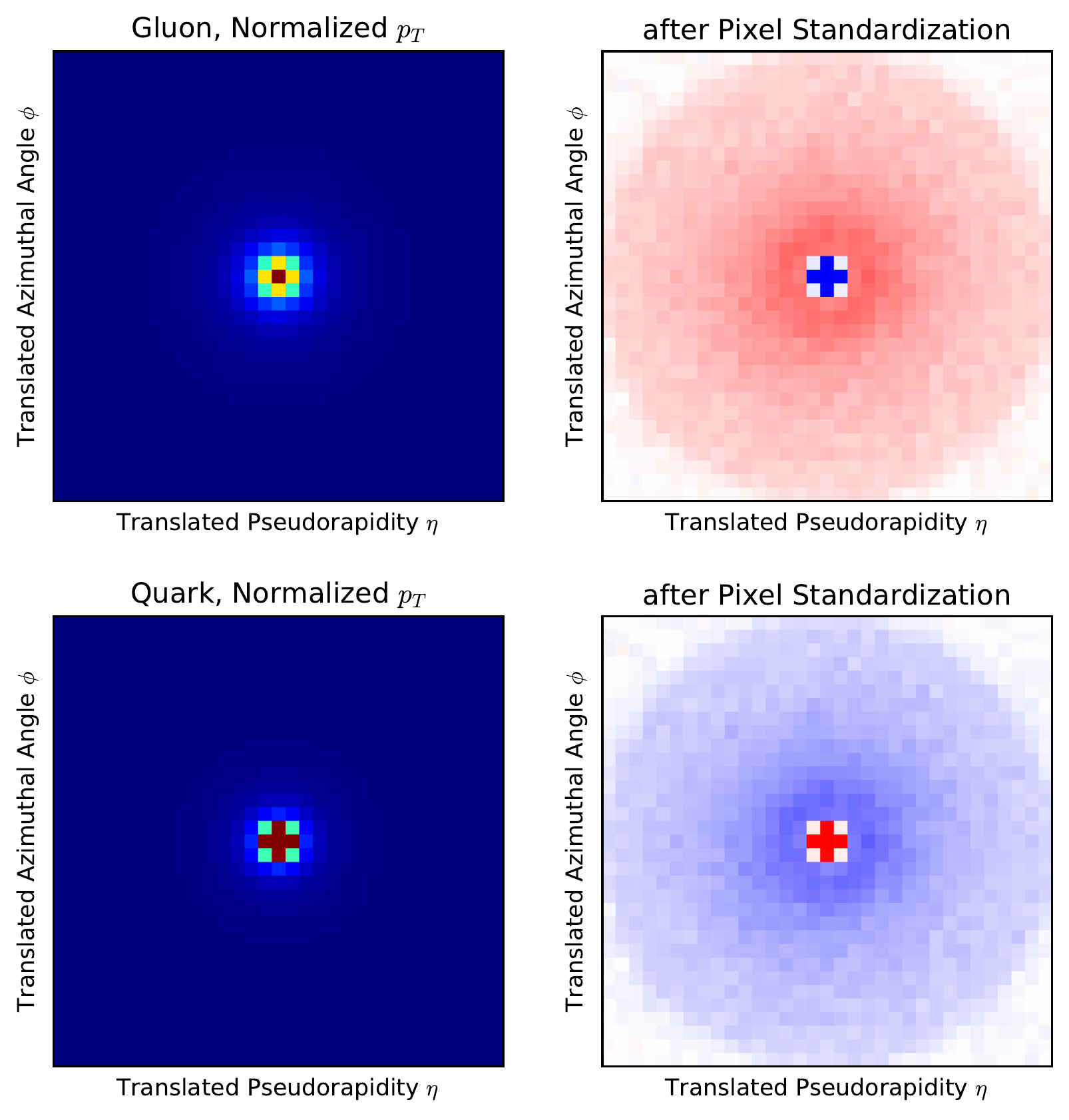}
\caption{The average jet images for 200 GeV Pythia gluon jets (top) and quark jets (bottom) shown after normalization (left) and after the zero-centering and standardization (right). Different linear color scales are used to highlight the important features of each step. On the left the quark jets have more intensity in the five core pixels whereas the gluon jets are wider. On the right, the standardization procedure illustrates that quark jets are narrower and emphasizes the softer outer radiation.}
\label{fig:preprocessing}
\end{figure}

Steps~\ref{step:zc} and \ref{step:stand} are additional pre-processing steps, not used in~\cite{deOliveira:2015xxd}. These new steps are often used in machine learning applications. We apply them here to put all the input pixels on an equal footing and allow the algorithm to more efficiently learn features to discriminate between classes. An improvement in performance was found after performing these additional pre-processing steps. Figure~\ref{fig:preprocessing} shows the average centered jet images for quark jets and gluon jets before and after these two new pre-processing steps. 

In addition, we implemented another useful pre-processeing step, called {\it data augmentation}~\cite{simard:2003}: for each jet image, its three reflections about the horizontal and/or vertical axes as well as its four translations vertically or horizontally by 1 pixel were added to the dataset. These transformations enforce the discrete symmetries of the configuration and make the model robust to possible centering issues. Though our samples were not statistics limited (generating more events is relatively cheap), such an approach may be helpful in circumstances where one cannot generate more samples (e.g. with real data or with full simulation). It also helps the network to learn invariance under certain symmetries. An additional possible data augmentation would be to include additional soft particles. These could represent uncertainties in underlying event modeling or pileup and would make the model more robust to noise. We did not explore this additional augmentation step in this work.

\subsection{Network architecture}
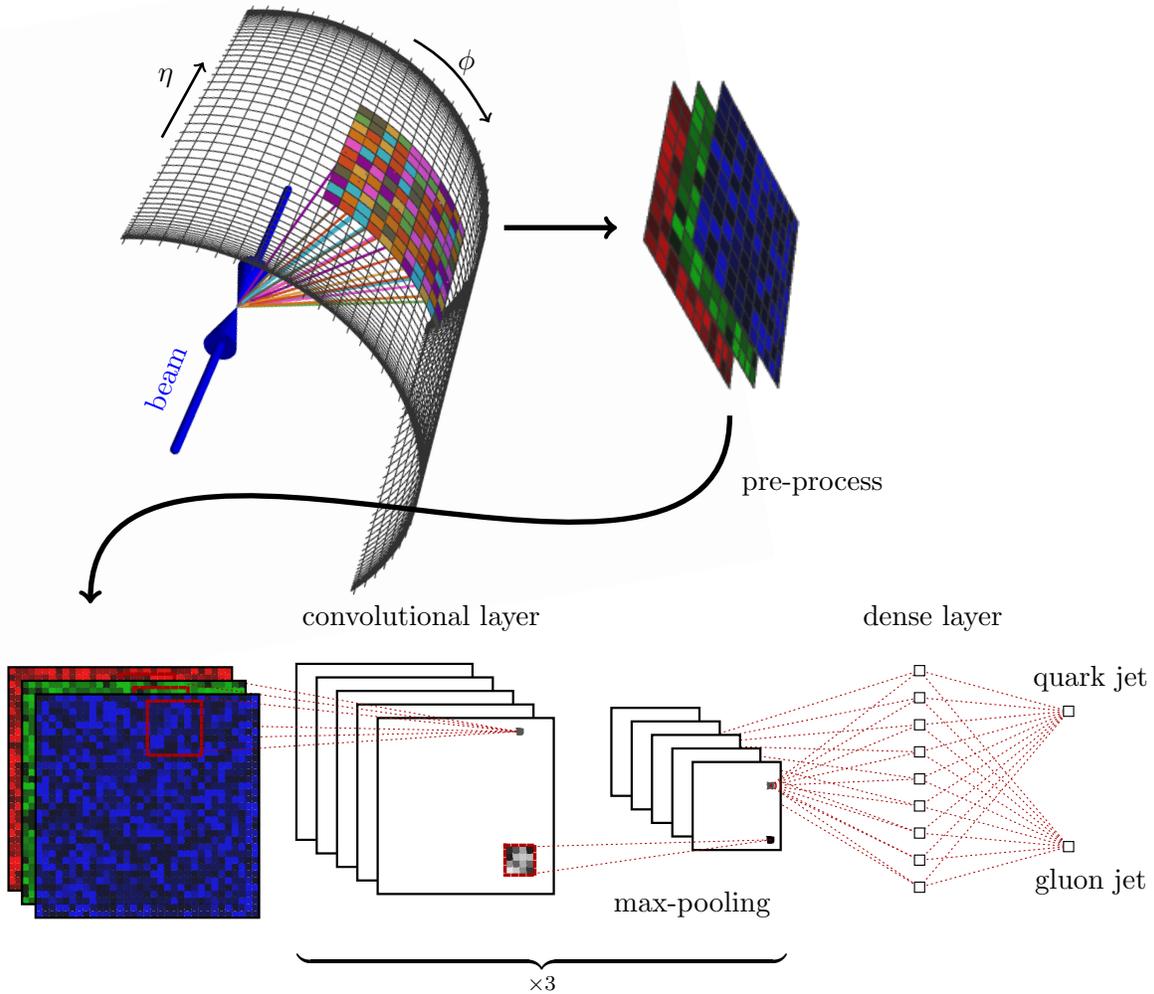
\begin{figure}[t]
\hspace{-2cm}
{{
\begin{tikzpicture}
\node at (-3.8,-2.3) [rotate=10] {\includegraphics[width=0.7\columnwidth, trim = {0 0 0 30mm}]{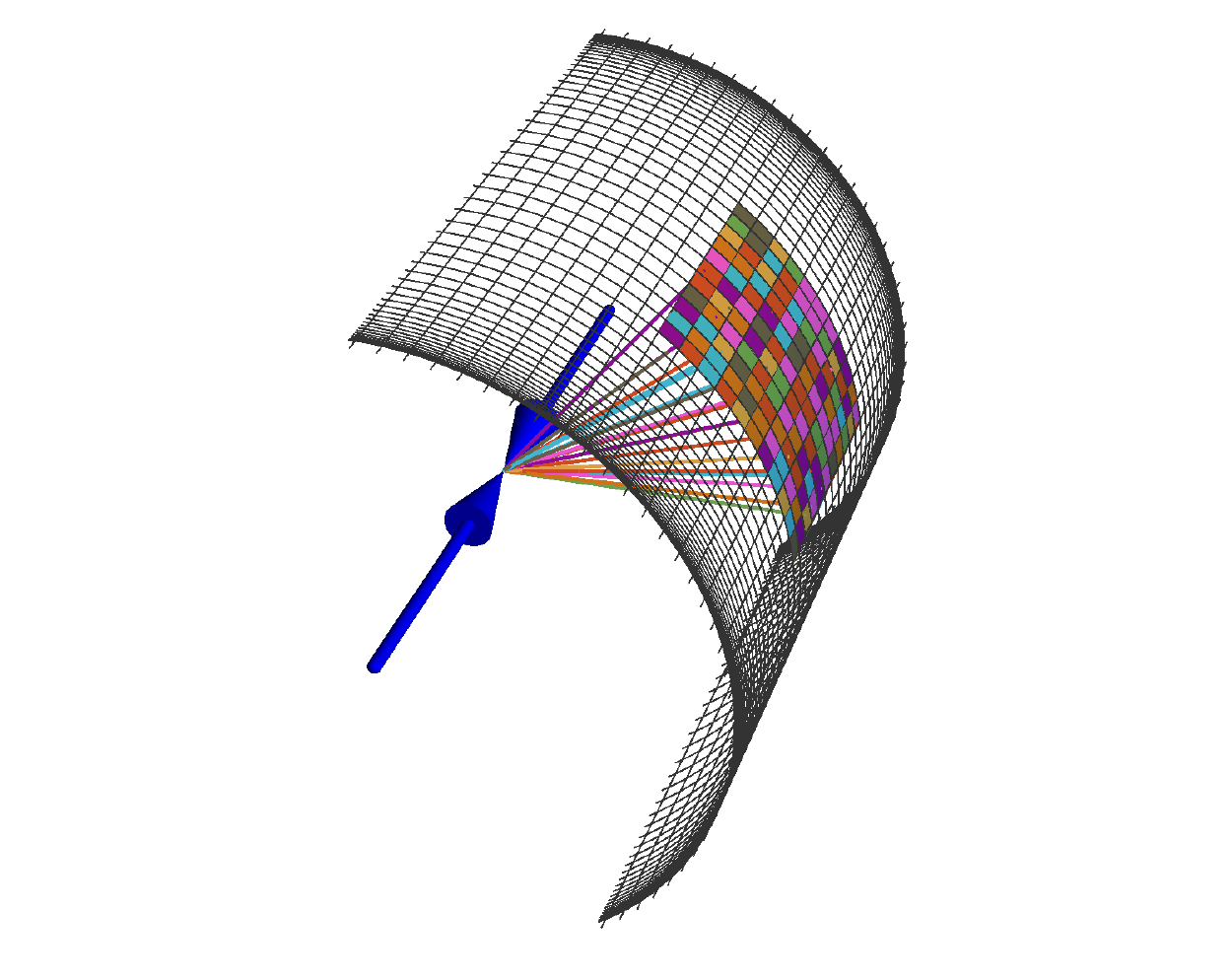}};
\draw [->,line width=2] (-1.5,0) -- (0,0);
\node[rotate=0] at (-6,2) {$\eta$};
\draw [->,line width=1] (-6.05,1.2) to (-5.5,2.2);
\node[rotate=0] at (-2,2.2) {$\phi$};
\draw [->,line width=1] (-2.7,2.5) to [out=-30, in = 120] (-1.7,1.4);
\node[rotate=68,,blue] at (-6,-2) {beam};
\node [above] at (2.6,-3.7) {pre-process};
\node at (1,0) {\includegraphics[width=0.5\columnwidth]{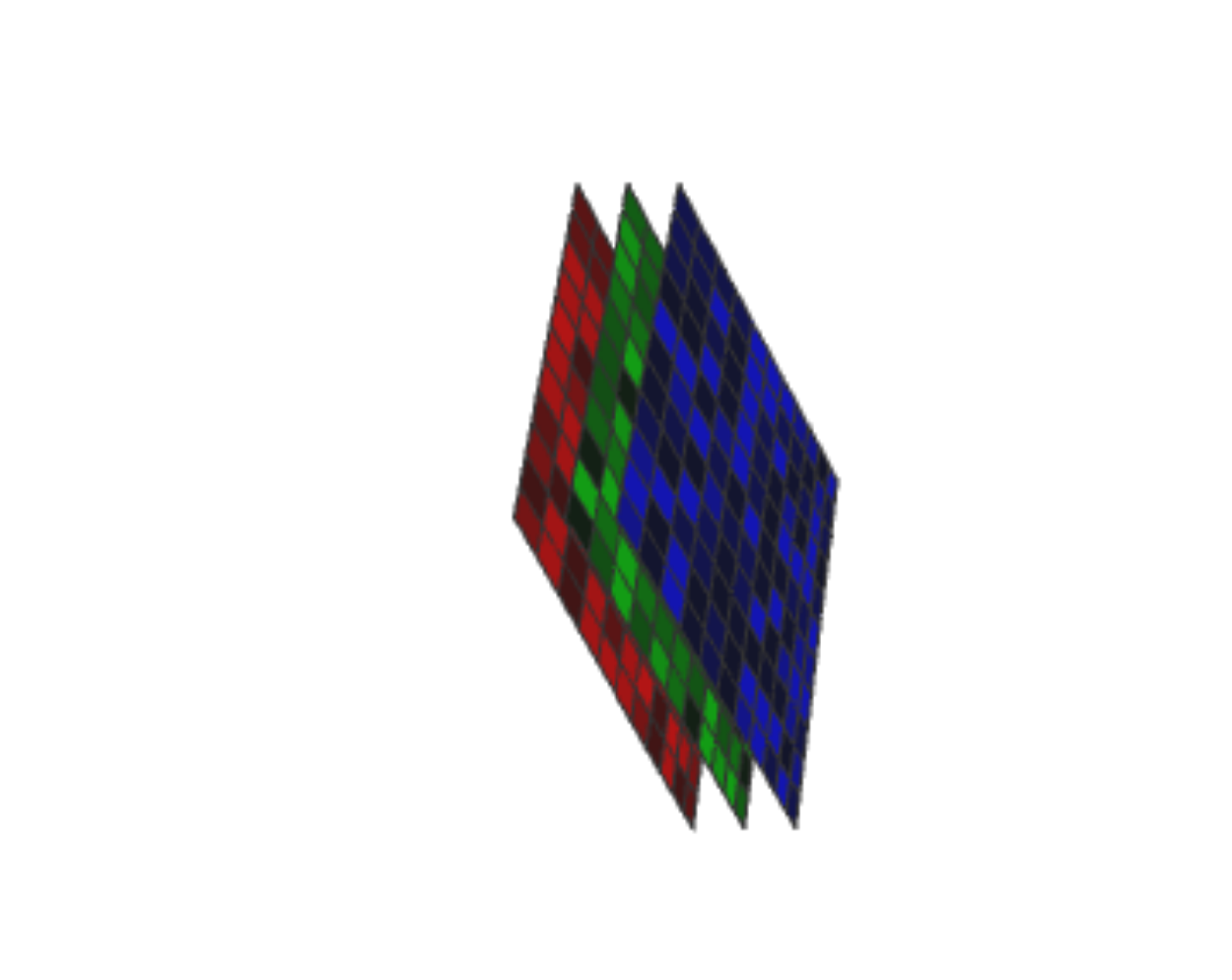}};
\draw[->,line width=2] (1.5,-2.5) to [out=-90,in=90] (-7,-5);
\node at (-1,-7.5)  {\includegraphics[width=0.95\columnwidth]{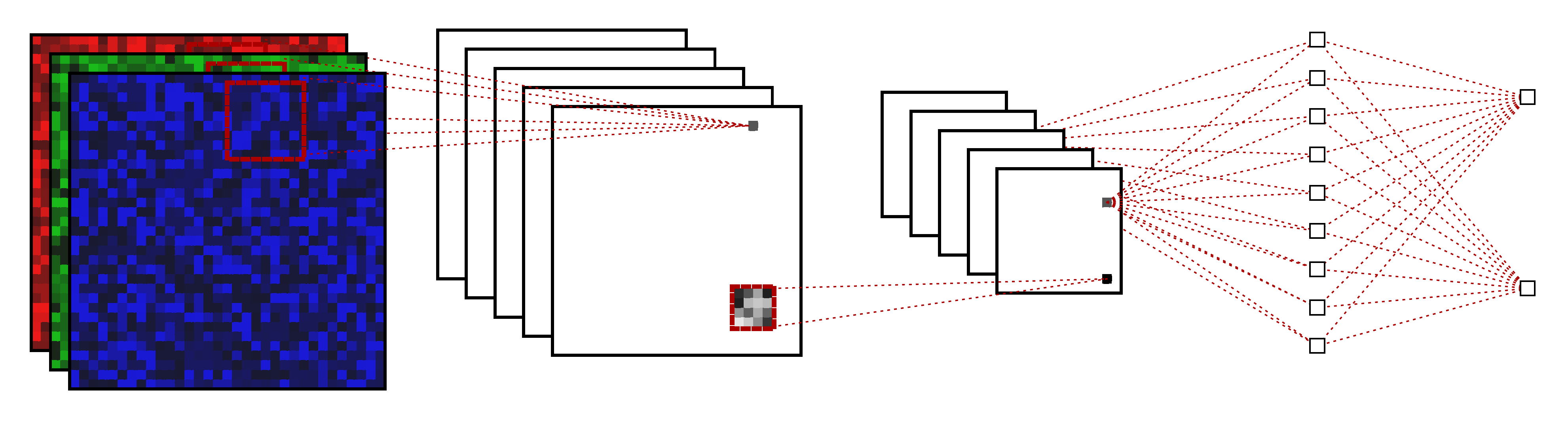}};
\node [above] at (-2.6,-5.5) {convolutional layer};
\node [below] at (1.0,-8.7) {max-pooling};
\node [above] at (4.2,-5.5) {dense layer};
\node [above] at (6.3,-6.3) {quark jet};
\node [above] at (6.3,-9) {gluon jet};
\node [below] at (-1,-9.5) {$\underbrace{\hspace{6.5cm}}_{ \times 3}$};
\end{tikzpicture}
}}
\caption{An illustration of the deep convolutional neural network architecture. The first layer is the input jet image, followed by three convolutional layers, a dense layer and an output layer. }
\label{fig:netarch}
\end{figure}

The deep convolutional network architecture used in this study consisted of three iterations of a convolutional layer with a ReLU activation and a maxpooling layer, all followed by a dense layer with a ReLU activation. To predict a binary classification between quarks and gluons, an output layer of two units with softmax activation is fully connected to the final dense hidden layer. An illustration of the architecture used is shown in Figure~\ref{fig:netarch}. The dropout rate was taken to be 0.25 after the first convolutional layer and $0.5$ for the remaining layers, with spatial dropout (drop entires feature maps) used in the convolutional layers. Each convolutional layer consisted of 64 filters, with filter sizes of $8\times 8$, $4\times 4$, and $4\times 4$, respectively. The maxpooling layers performed a $2\times 2$ down-sampling with a stride length of 2. The dense layer consisted of 128 units. 

All neural network architecture training was performed with the Python deep learning libraries Keras~\cite{keras} and Theano~\cite{theano} on NVidia Tesla K40 and K80 GPUs using the NVidia CUDA platform. The data consisted of the 100k jet images per $p_T$-bin, partitioned into 90k training images and 10k test images. An additional $10\%$ of the training images are randomly withheld as validation data during training of the model for the purposes of hyperparameter optimization. He-uniform initialization~\cite{heuniform} was used to initialize the model weights. The network was trained using the Adam algorithm~\cite{adam} using categorical cross-entropy as a loss function with a batch size of 128 over 50 epochs and an early-stopping patience of 2 to 5 epochs.

Only moderate optimization of the network architecture and minimal hyperparameter-tuning were performed in this study. This optimization included exploration of different optimizers (Adam, Adadelta, RMSprop), filter sizes, number of filters, activation functions (ReLU, tanh), and regularization (dropout, $L_2$-regularization), though this exploration was not exhaustive. Further systematic exploration of the space of architectures and hyperparameter values, such as with Bayesian optimization using Spearmint~\cite{spearmint}, might increase the performance of the deep neural network. 

\subsection{Jet images in color \label{sec:colorn}}
All implementations of the jet images machine learning approach that we know of take as the input image a grid where the input layer contains the pre-processed energy or transverse momentum in a particular angular region. This can be thought of as a grayscale image, with only intensity in each pixel and all color information discarded. In computer vision one can do better by training on color images, with red, green and blue intensities treated as separate input layers, also known as channels. Thus, it is natural to try to apply some methods for processing color to physics applications.

For particle physics, there are many ways the calorimeter deposits can be partitioned. One could try to identify the actual particles: have one channel for protons, one for neutrons, one for electrons, one for $\pi^+$ particles, one for $K_L$'s, etc. Although it is not yet possible to completely separate every type of metastable particle, advances in experimental techniques, such as particle flow~\cite{CMS:2009nxa}, indicate that this may not be too unrealistic. However, it is also not clear that having 15 color channels would help and training with so many input channels would be much slower. There are many options for a smaller set of channels. For example, one could consider one channel for hadrons and one for leptons, or channels for positively charged, neutral and negatively charged particles. To be concrete, in this study we take three input channels:
\begin{enumerate}
\item[] {\color{darkred}{\tt red} = transverse momenta of charged particles}
\item[] {\color{darkgreen}{\tt green} =  the transverse momenta of neutral particles}
\item[] {\color{darkblue}{\tt blue} = charged particle multiplicity}
\end{enumerate}
Each of these observables is evaluated on each image pixel. All channels of the image undergo the following pre-processing: the images are normalized such that the sum of the red and green channels is one; the zero centering and standardization are done for each pixel in each channel according to $I_{ij}^{(k)} \to (I_{ij}^{(k)} - \mu_{ij}^{(k)})/(\sigma_{ij}^{(k)} + r)$. Here, $I_{ij}^{(k)}$ is the intensity of pixel $ij$ in channel $k$ of an image, and $\mu_{ij}^{(k)}$ and $\sigma_{ij}^{(k)}$ are the respective mean and standard deviation of pixel $ij$ in channel $k$ in the training data.

The network architecture is designed to respect the overlay of the different color images. That is, every image channel feeds into the same units in the network and the weights from each channel are allowed to vary independently. In other words, in the first convolutional layer instead of using an $8\times 8$ filter with 64 weights, we use an $8\times 8\times 3$ filter with 192 weights. The number of units in the first layer is the same as with grayscale inputs, and the rest of the network architecture is the same with or without colored jet image inputs.

\section{Fisher jets and a look inside the networks \label{sec:fisher}}
One potentially useful output of a convolutional neural network (other than the network itself) is the learned filters. These filters can display features of the image trained at different levels of resolution. In facial recognition applications, for example, one filter might pick up on noses, another on eyes, and so on.

If the entire network just had one unit, connected to all the inputs by different weights, these weights themselves would form an image and the network would act by taking the inner product of the input image and the weight image. Such a procedure, with an appropriate loss function, is equivalent to Fisher's Linear Discriminant (FLD), which was applied to jet images in~\cite{Cogan:2014oua}.   This weight image is the simplest example of a filter, and so we first find the FLD for our quark/gluon samples using the Python machine learning package scikit-learn~\cite{sklearn}. To be more precise, the FLD method determines a Fisher jet image $F$ that maximizes a separation by the discriminant $D$ defined on a (grayscale) jet image $I$ by:
\begin{equation}
D[I] = \sum_{ij} F_{ij} I_{ij}.
\end{equation}

\begin{figure}
\centering
\includegraphics[scale=0.503]{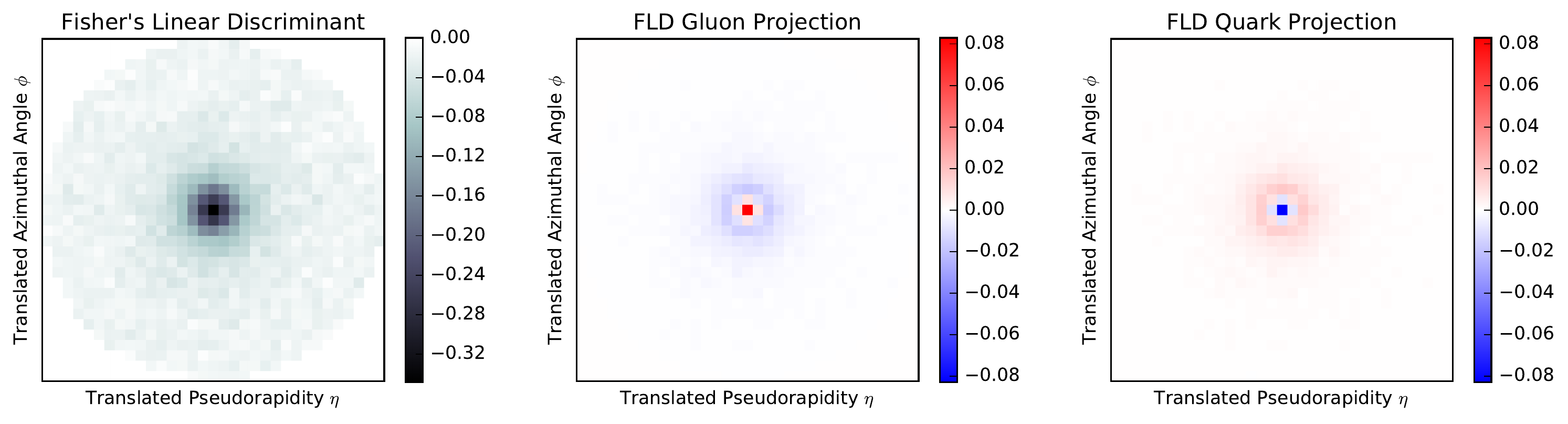}
\caption{(left) The weight vector of Fisher's linear discriminant shown as a Fisher jet image, trained on 200 GeV {Pythia} jets with the additional log transformation step included in the pre-processing. The average standardized 200 GeV quark jet image (middle) and gluon jet image (right) shown after being projected with the Fisher jet.}
\label{fig:fisherjet}
\end{figure}

Without applying the zero-centering and standardization in the pre-processing, regularization must be applied to the FLD as in~\cite{Cogan:2014oua} to prevent overfitting and reduce the sensitivity of the Fisher jet to the noisy outer bits of the jet images. However, after the full pre-processing of the jet images, no regularization is necessary to arrive at a sensible Fisher jet. An additional pre-processing step of a log transformation $I_{ij} \to \log(1 + I_{ij}/r')$ with $r' = 10^{-3}$ before the zero-centering was found to significantly improve the FLD performance, though this log transformation was not found to be necessary for training the deep neural networks. The Fisher jet resulting from the FLD analysis with the additional log transformation on \SI{200}{GeV} jet images and its pixel-wise product with the average quark and gluon jet images are shown in Figure~\ref{fig:fisherjet}. The discriminating performance of the FLD is shown and compared to various jet variables and the deep convolutional network in Table~\ref{tab:effs} and Figure~\ref{fig:CNN_baselines}.

\begin{figure}
\centering
\includegraphics[scale=0.75]{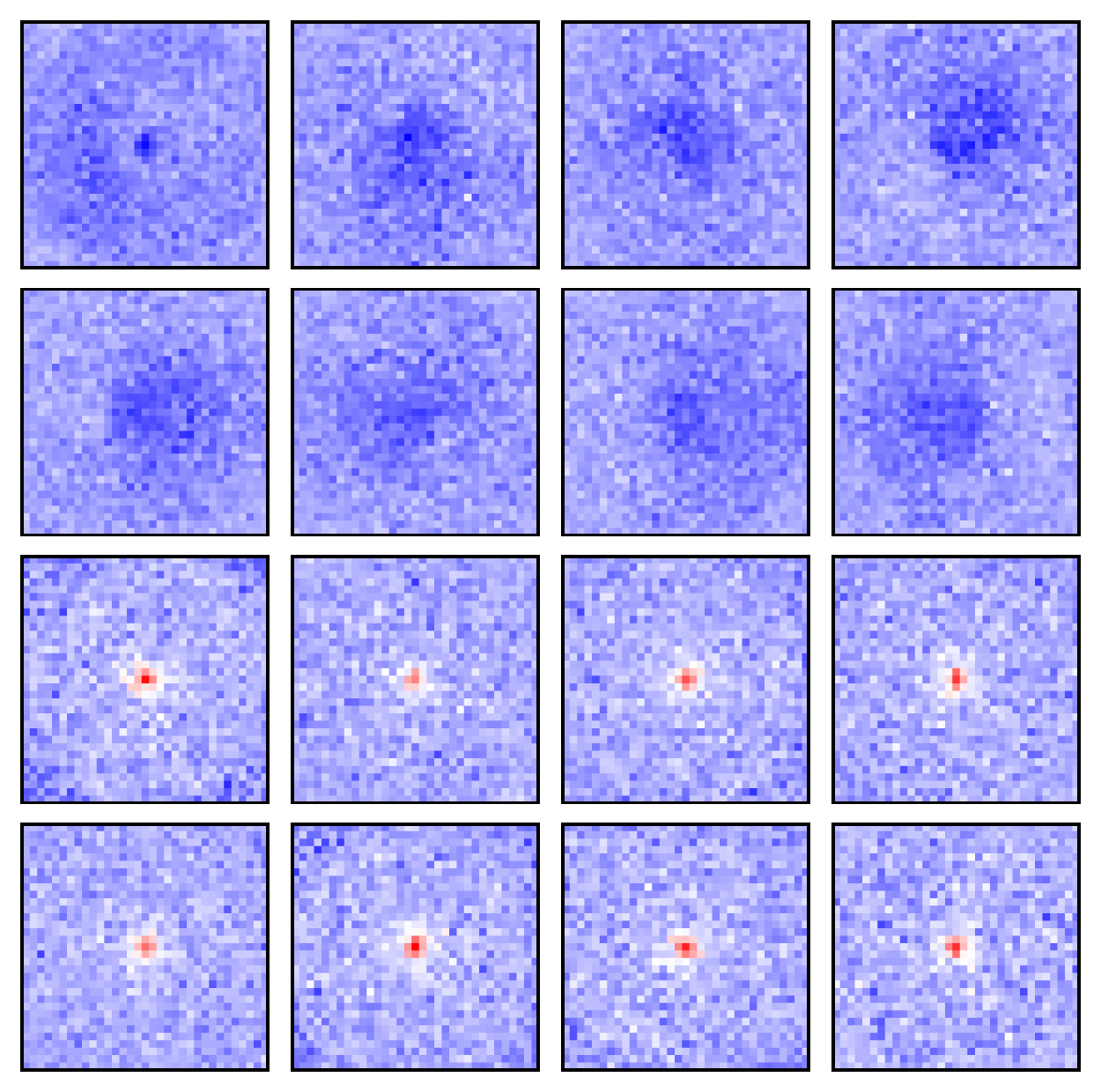}
\caption{The 16 sets of $33\times 33$ weights learned by the shallow dense neural network with 16 units on pre-processed 200 GeV {Pythia} jets. Red and blue indicate positive and negative intensity, respectively. The images are ordered from smallest to largest by the sum of their weights. The learned observables fall into two categories: observables sensitive to the large-scale jet shape and observables sensitive to the $p_T$ near the center of the image. The first eight observables pick up on geometric moments of the jet shape, including an annular structure in the first image and several left/right asymmetric structures.}
\label{fig:shallowDNNfilters}
\end{figure}

The next step up from the FLD is a shallow dense neural network, consisting of one dense layer and two output sigmoid units. This is a natural generalization of the FLD analysis. The additional units allow the network to learn more discriminating features. We trained a single-layer network consisting of a 16 fully-connected units with a ReLU activation, an L2-regularization parameter of $10^{-7}$ on all the weights and bias terms, and a dropout rate of 0.25. The log transformation is applied to the inputs, as it was found to increase the performance of shallow networks. The 16 sets of $33\times 33$ weights corresponding to the observables that this model learns are shown in Figure~\ref{fig:shallowDNNfilters}. The observables generally fall into two classes: those sensitive to large geometric moments of the jet shape and those that are sensititve to the transverse momenta close to the core of the jet. The discriminating performance of this shallow dense network is compared to jet variables and the convolutional network in Table~\ref{tab:effs}.

\section{Network performance \label{sec:performance}}
A thorough study of physics-motivated jet variables for quark/gluon discrimination was performed in~\cite{Gallicchio:2012ez}, where continuous shape variables such as the jet girth and two-point moment and discrete variables such as the charged particle multiplicity were considered. To compare the performance of the neural networks trained on jet images to that of physics-motivated variables, the following five {jet observables} were considered:
\begin{itemize}
\item Girth: $\sum_{i} p_{T}^i r_i/ p_T^\text{jet}$, where $r_i = \sqrt{\Delta\phi_i^2 + \Delta\eta_i^2}$. The sum is taken over the pixels in the image to account for the discretization of the detector.
\item Charged Particle Multiplicity (CPM): The number of charged final-state particles in the jet. We did not apply any pixelation or detector simulation to this observable. 
\item Two-Point Moment~\cite{Larkoski:2013}: $\sum_i \sum_j p_T^i p_T^j r_i^\beta / (p_T^\text{jet})^2$, where the value $\beta = 0.2$ is used. The sum is taken over the pixels in the image to account for the discretization of the detector.
\item $x_\text{max}$~\cite{Kucuk:2016}: The highest fraction of the total $p_T$ contained in a single pixel.
\item $N_\text{95}$~\cite{Pumplin:1991kc}: The minimum number of pixels which contain 95\% of the total $p_T$ of the jet.
\end{itemize}

The jet variable $N_\text{95}$ was introduced (as $N_{90}$) for quark/gluon discrimination by~\cite{Pumplin:1991kc} in 1991, where a framework very similar to jet images was also introduced. In~\cite{Pumplin:1991kc}, $N_{95}$ was found to be a single variable which outperformed neural networks at quark/gluon discrimination at that time, and we find that it is the physics-motivated observable with the best performance in several cases. Optimizing over the fraction of the jet $p_T$ to consider, $N_{95}$ performs better than $N_{90}$ for the samples considered in this study. Deep convolutional nets notably outperform this variable, indicating an advantage from the deep learning with jet images approach over previous uses of neural networks for quark/gluon discrimination.

We measure the discrimination power of an observable by the lowest achievable gluon acceptance efficiency $\varepsilon_g$ at a given quark acceptance efficiency $\varepsilon_q$. The performance can be visualized through a receiver operating characteristic (ROC) curve, which plots $1-\varepsilon_g$ as a function of $\varepsilon_q$. An alternative visualization is the significance improvement characteristic (SIC) $\varepsilon_q/\sqrt{\varepsilon_g}$. A SIC curve has the advantage of being closely connected to the improvement in signal over background discrimination power in a collider physics application, and also exhibits a nontrivial maximum (at some $\varepsilon_q$) which gives an unbiased measure of the relative performance of different discriminants~\cite{Gallicchio:2010dq}. 

\begin{figure}
\centering
\begin{tikzpicture}
\node at (0,0) {\includegraphics[width=0.5\columnwidth]{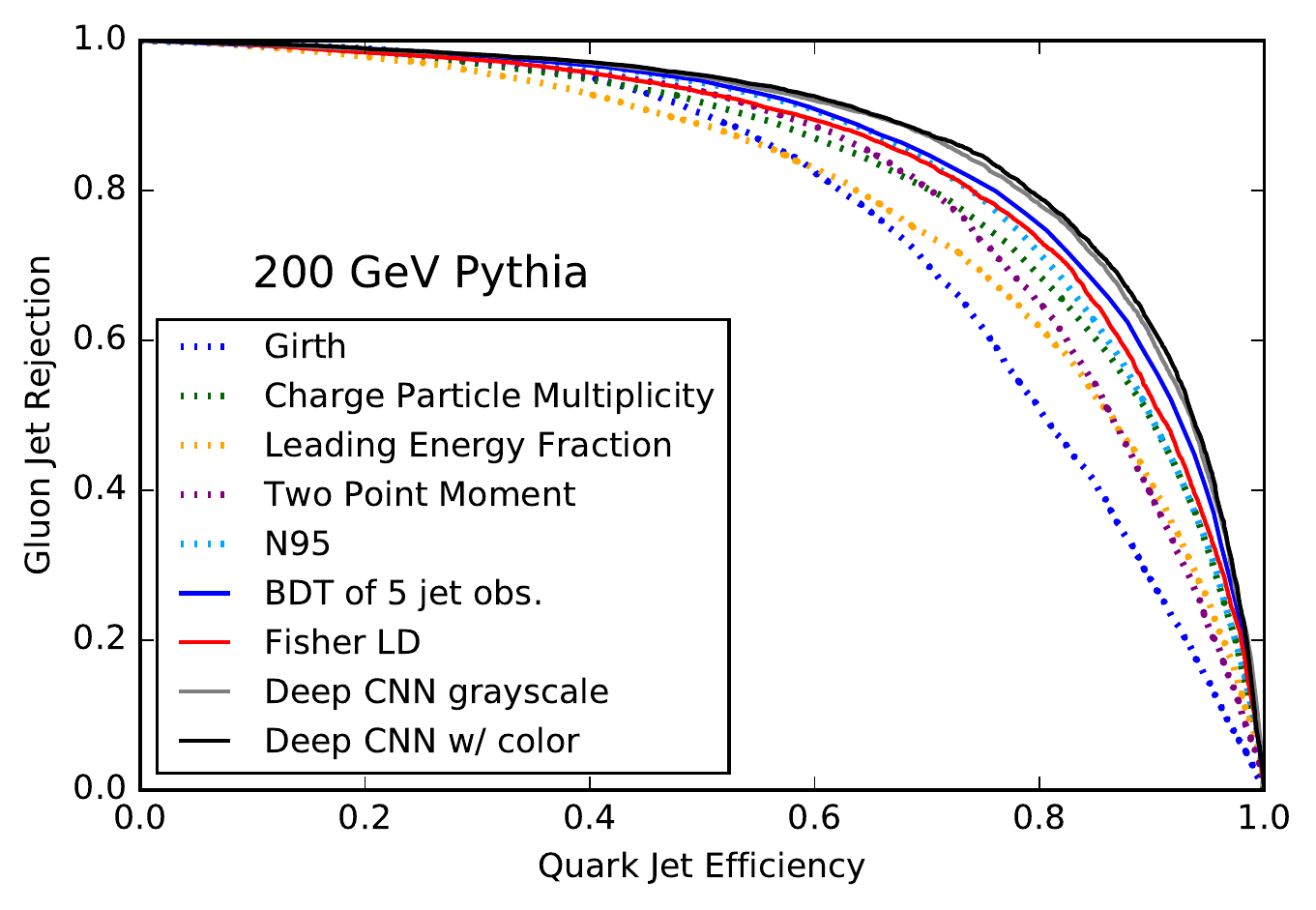}};
\node at (7.8,0) {\includegraphics[width=0.5\columnwidth]{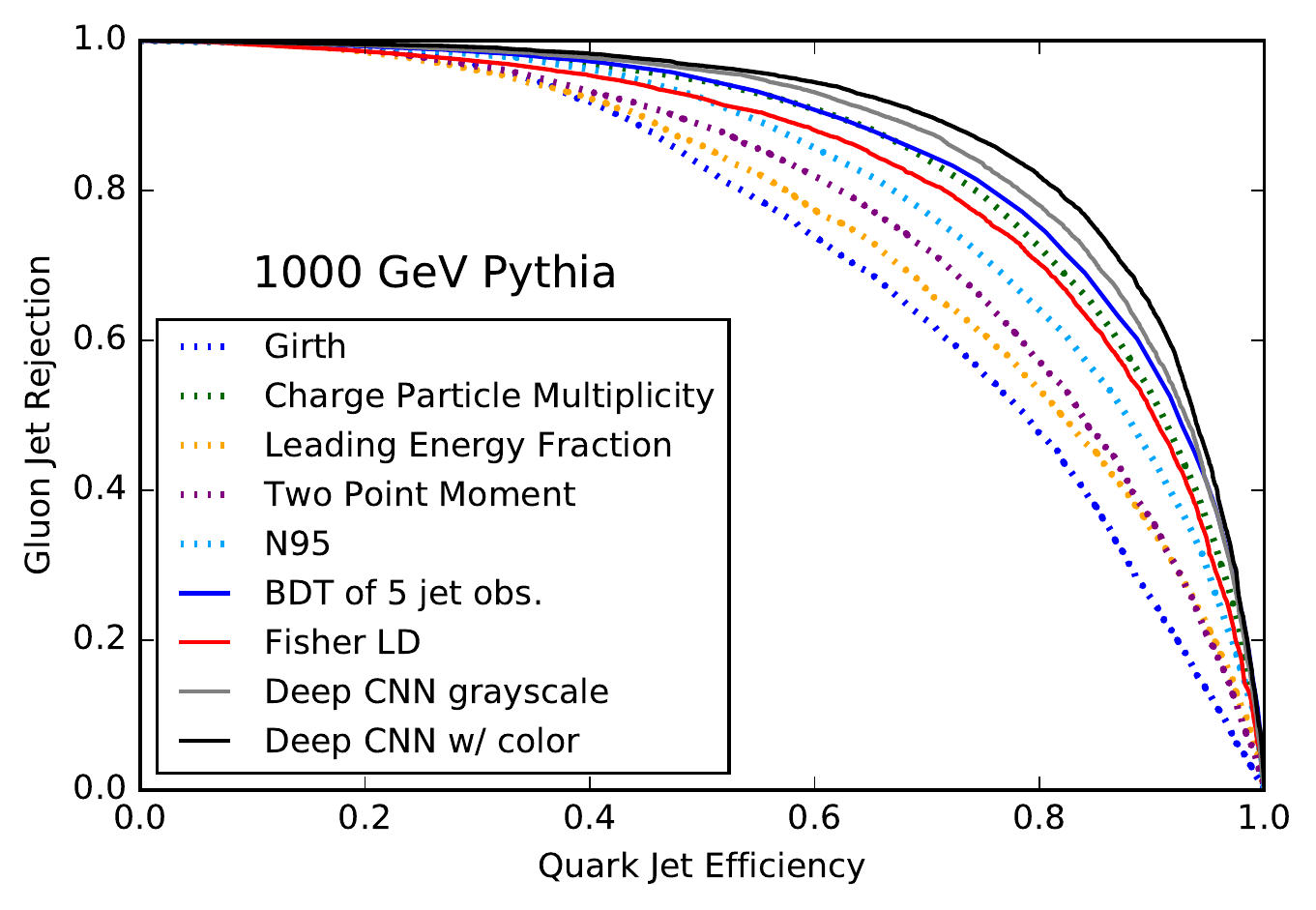}};
\end{tikzpicture}
\begin{tikzpicture}
\node at (0,0) {\includegraphics[width=0.5\columnwidth]{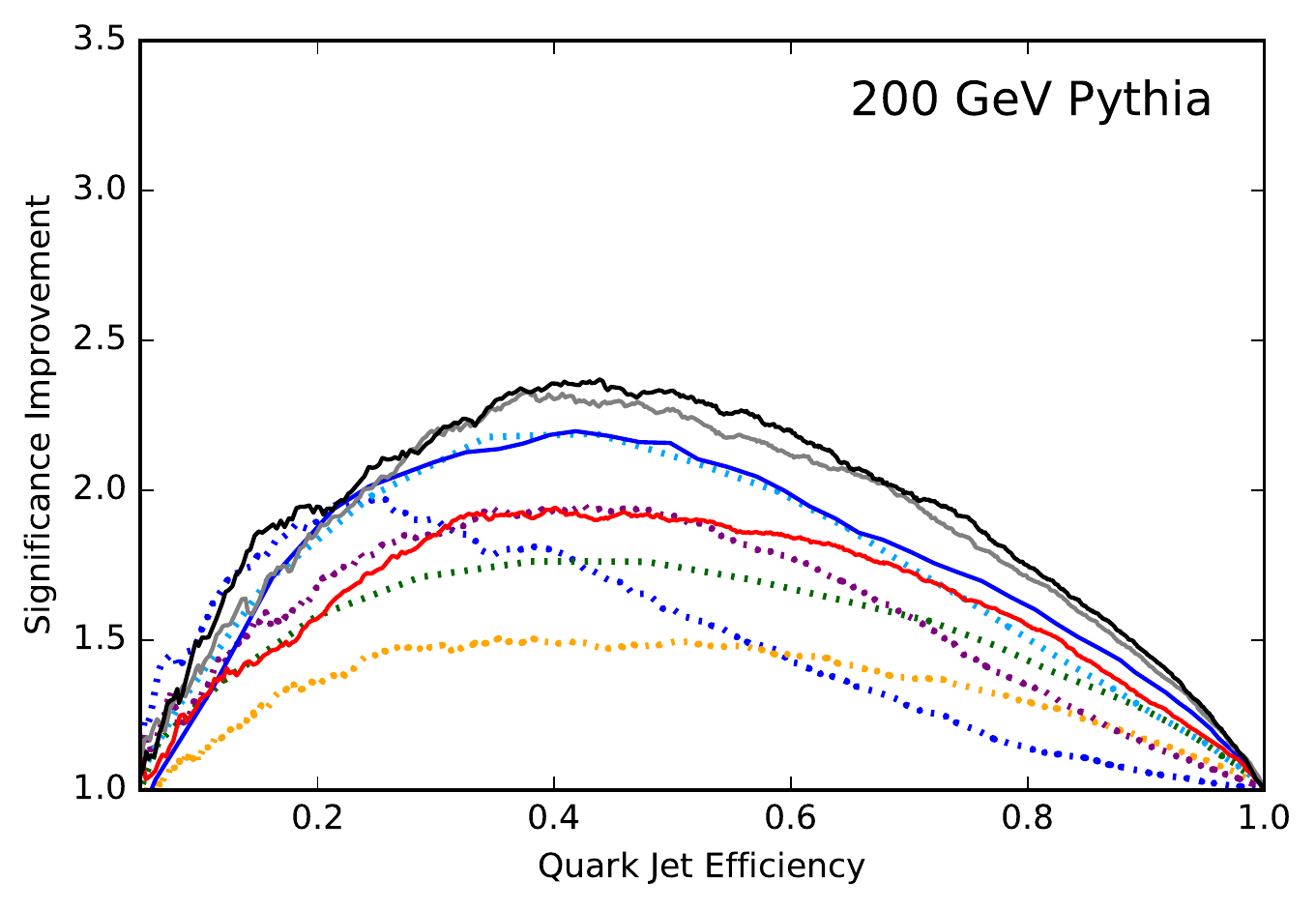}};
\node at (7.8,0) {\includegraphics[width=0.5\columnwidth]{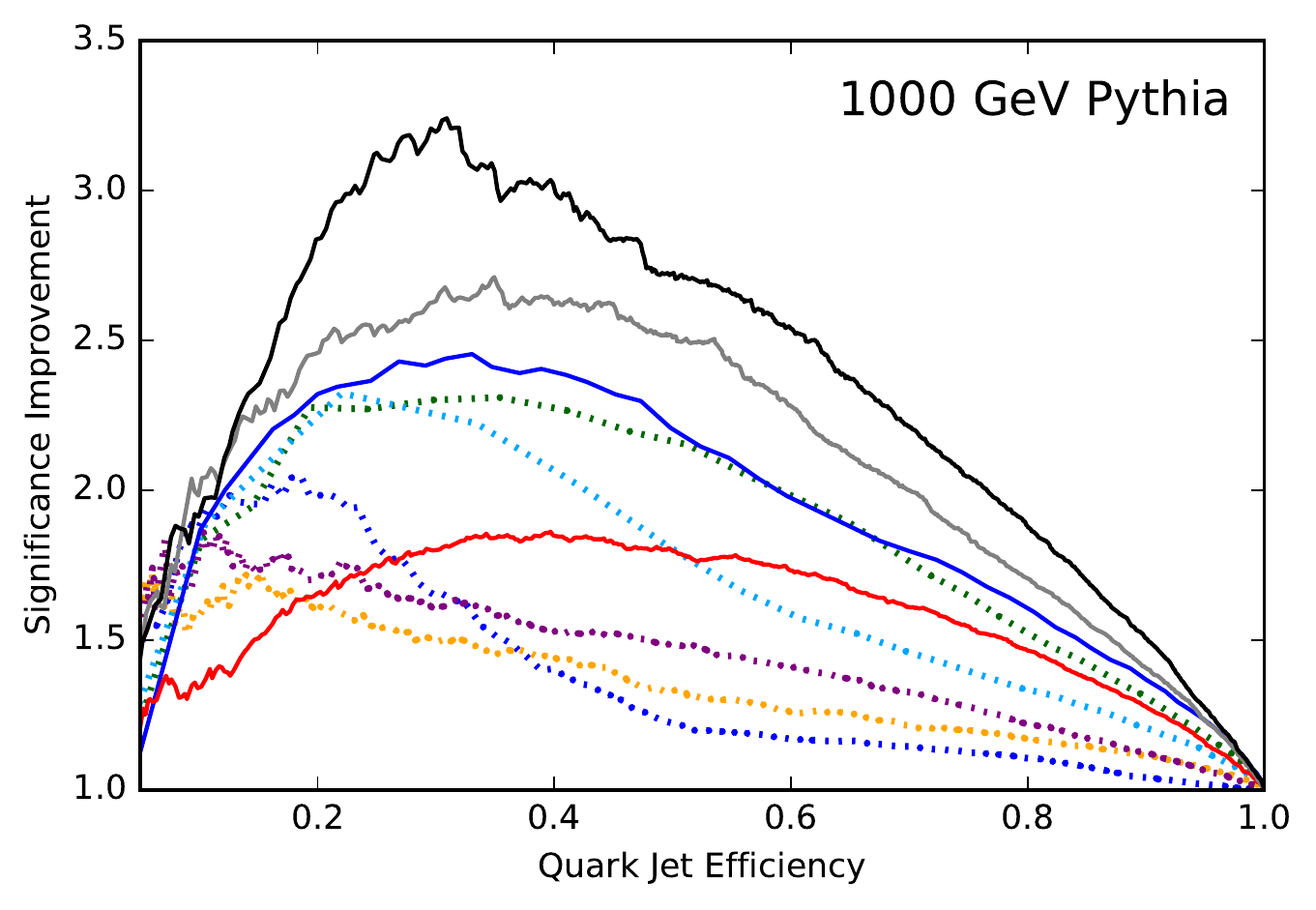}};
\end{tikzpicture}
\caption{\label{fig:CNN_baselines}(top) ROC and (bottom) SIC curves of the FLD and the deep convolutional network trained on (left) \SI{200}{GeV} and (right) \SI{1000}{GeV} {Pythia} jet images with and without color compared to baseline jet observables and a BDT of the five jet observables.}
\end{figure}

The ROC and SIC curves of the jet variables and the deep convolutional network on \SI{200}{GeV} and \SI{1000}{GeV} Pythia jets are shown in Figure~\ref{fig:CNN_baselines}. The quark jet classificiation efficiency at 50\% quark jet classification efficiency for each of the jet variables and the CNN are listed in Table~\ref{tab:effs}. To combine the jet variables into more sophisticated discriminants, a boosted decision tree (BDT) is implemented with scikit-learn. The convolutional network outperforms the traditional variables and matches or exceeds the performance of the BDT of all of the jet variables. The performance of the networks trained on images with and without color is shown in Figure~\ref{fig:color}.

\subsection{Colored jet images\label{sec:color}}
\begin{figure}[t]
\centering
\includegraphics[scale=0.75]{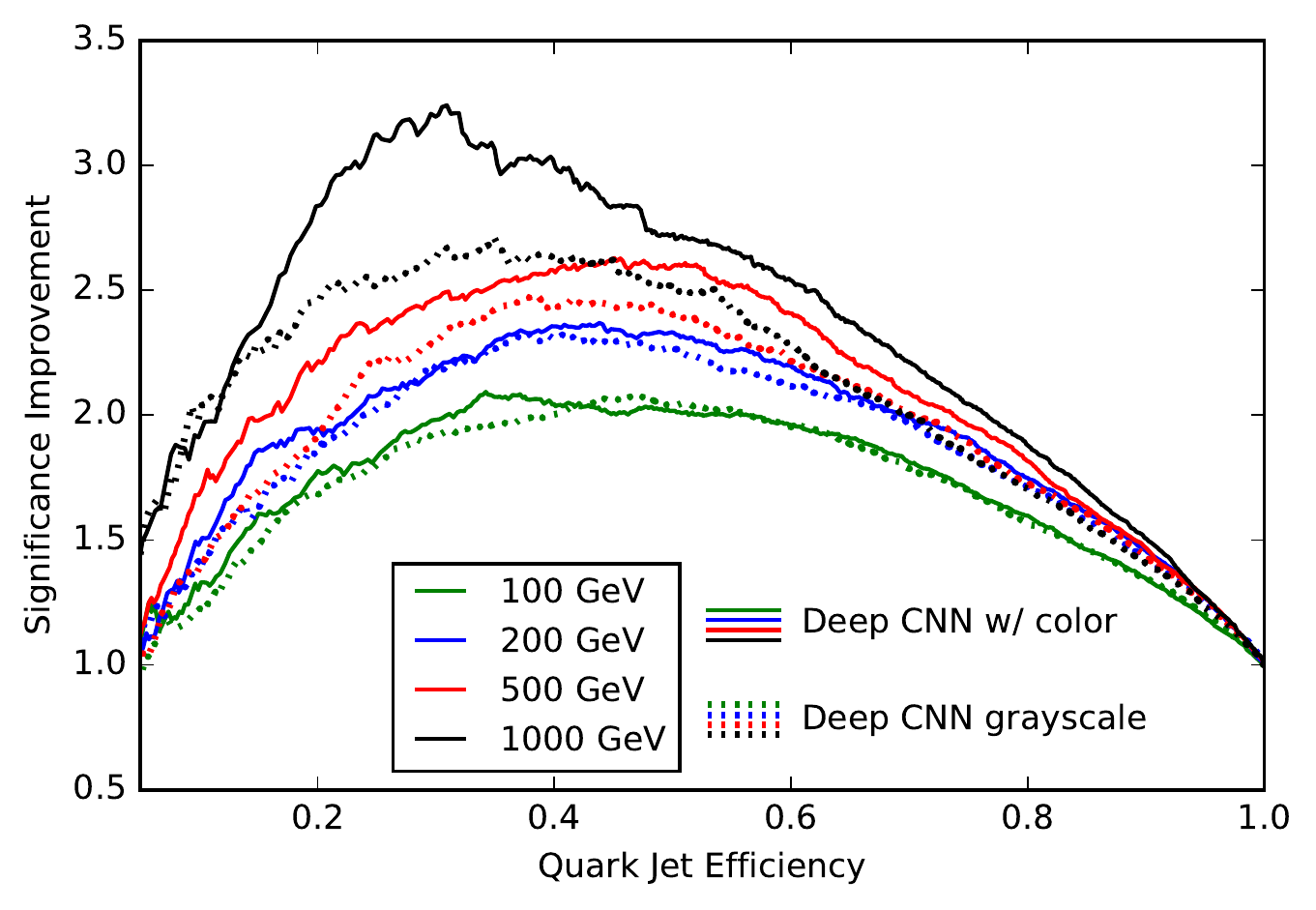}
\caption{\label{fig:color}SIC curve of deep convolutional network performance on {Pythia} jets with color (solid) and without color (dotted). The introduction of color becomes more helpful at higher energies, with the largest improvement on the \SI{1000}{GeV} jets.}
\end{figure}

The benchmarks in the previous section were compared to the jet images with and without color, where the three color channels correspond to separating out the charge and multiplicity information as described in Section~\ref{sec:colorn}. Figure~\ref{fig:color} shows the SIC curves of the neural network performances with and without color on {Pythia} jet images. For the \SI{100}{GeV} and \SI{200}{GeV} images, only small changes in the network performance were observed by adding in color of this form. For the \SI{500}{GeV} and \SI{1000}{GeV} jet images, performance increases were consistently observed by adding color to the jet images. From Table~\ref{tab:effs}, one sees that the improvement at high $p_T$ from using multiple channels occurs with both Pythia and Herwig samples.  

Alternative definitions of color, including positive, negative, and neutral $p_T$-channels, were considered and the present color definition was found to facilitate the best network performance within the considered architectures.

\subsection{Merge layers \label{sec:merge}}
One natural question about the neural network is whether it has learned information equivalent to basic jet variables such as CPM. To approach this question, jet observables were included as additional inputs to the model through a Keras merge layer. The jet observables were fed into a two-layer dense network with 128 units per layer. The output of the second dense layer was then merged with the output of the third convolutional layer in the deep CNN, with the remainder of the CNN unchanged. If one observes a significant improvement in performance when the network is given access to the jet variable, it would indicate that the network did not learn information equivalent to that jet variable.  

\begin{figure}[b]
\centering
\includegraphics[scale=0.75]{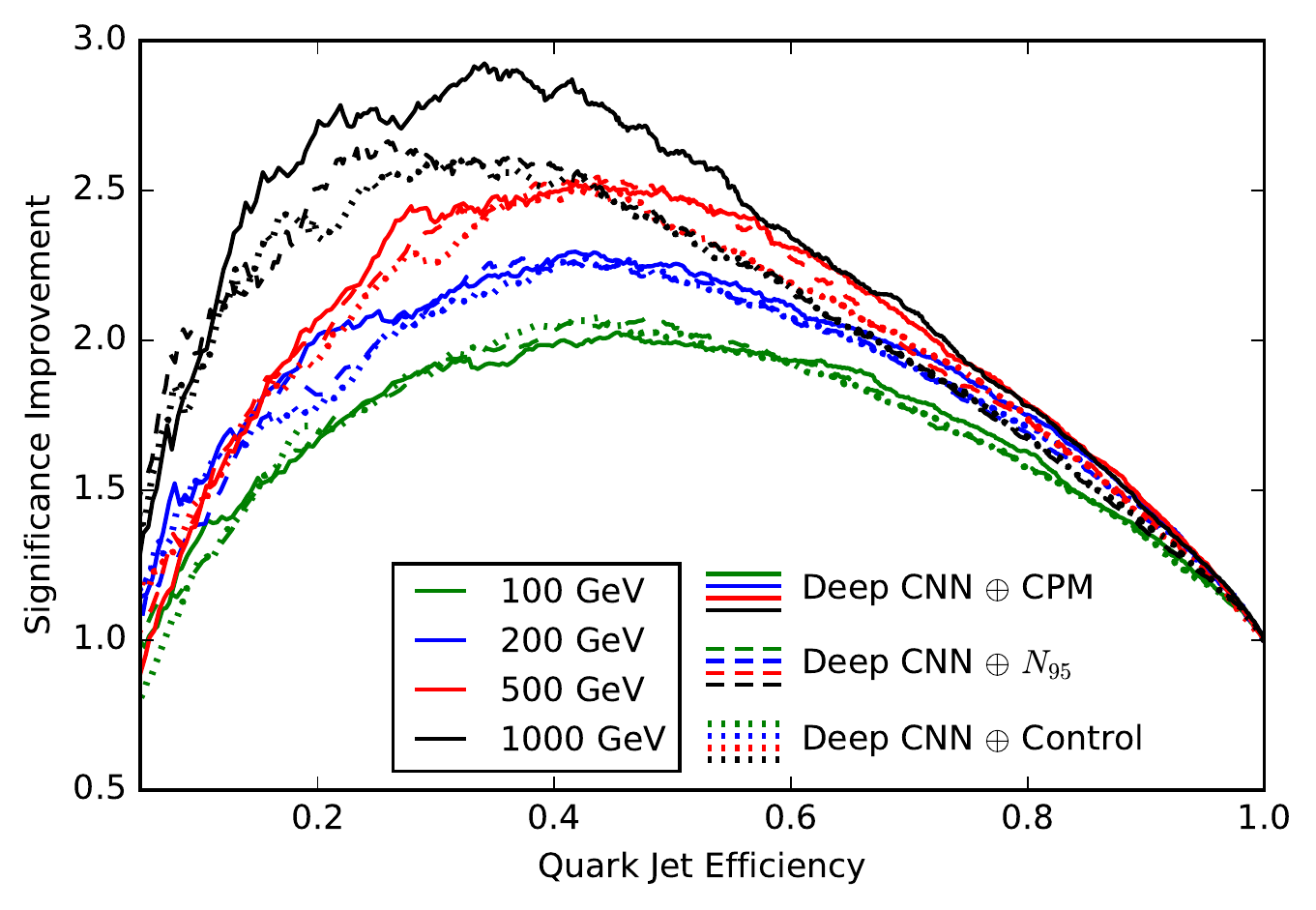}
\caption{\label{fig:merge}SIC curve of deep convolutional network performance on {Pythia} jets without color with additional inputs of CPM (solid), $N_{95}$ (dashed), and zero as a control (dotted). The spread in the SIC curves for models trained on  \SI{100}{GeV} and \SI{200}{GeV} jets is within the typical variation, with no clear improvement from the additional variables. For models trained on \SI{500}{GeV} and \SI{1000}{GeV} jets, a modest improvement was seen from the introduction of CPM, though not as large as the improvement from the introduction of color.}
\end{figure}

Figure~\ref{fig:merge} shows the SIC curves of the neural network performance on jets without color trained with additional inputs of zero, $N_{95}$, or CPM. Modest improvement in performance is found by adding CPM as an additional input at \SI{500}{GeV} and \SI{1000}{GeV}, whereas little or no improvement is found by adding $N_{95}$ as an additional input. This behavior is indication that the network is learning geometric observables which contain information similar to $N_{95}$.

\begin{table}
\centering
\begin{tabular}{|l|c|c|c|c|}
\hline
\bf Gluon Jet Efficiency (\%) at & \SI{1000}{GeV}& \SI{1000}{GeV} &\SI{200}{GeV} & \SI{200}{GeV} \\ 
\bf  50\% Quark Jet Acceptance & {Pythia}& {Herwig} & {Pythia} & {Herwig}\\
\hline
Deep CNN with Color & 3.4 & 18.6 & 4.6 & 16.5\\
Deep CNN without Color & 4.0 & 22.2 & 4.8 & 16.4\\
Shallow Dense Network & 6.0& 23.3 & 5.5& 18.4\\
Fisher's Linear Discriminant & 7.4 & 24.1 & 6.3 & 17.7\\ \hline
BDT of all jet variables & 5.2 & 21.0 & 5.2 & 16.4 \\ 
Girth $\times$ CPM &9.6  & 31.8 & 6.7 & 20.5 \\ \hline
$N_{95} $ & 7.4 & 26.9 & 6.1 & 19.6 \\
Charged Particle Multiplicity (CPM) & 5.7 & 20.4 & 7.4 & 20.4 \\
Two Point Moment & 11.3  & 26.8 & 6.9 & 17.4\\
Girth & 16.9 & 31.8 & 9.8 & 23.4 \\ 
Leading energy fraction $x_\text{max}$& 14.1 & 28.1 & 11.1 & 21.1 \\
\hline
\end{tabular}
\caption{Gluon efficiencies at 50\% quark acceptance for \SI{200}{GeV} and \SI{1000}{GeV} Pythia and Herwig jets using $33\times33$ images. Girth $\times$ CPM is the product of these two observables, as motivated in~\cite{Gallicchio:2012ez}.}
\label{tab:effs}
\end{table}

\begin{table}
\centering
\begin{tabular}{|l|c|c|c|c|}
\hline
\bf  Image Size & $13\times 13$ & $23\times 23$ & $33\times 33$ & $43\times 43$ \\
\hline
\bf Gluon Jet Efficiency (\%) at & \multirow{2}{*}{6.1}  & \multirow{2}{*}{5.7} & \multirow{2}{*}{5.5} & \multirow{2}{*}{5.7} \\
\bf 50\% Quark Jet Acceptance &&&&\\
\hline
\end{tabular}
\caption{Gluon jet efficiencies at 50\% quark jet acceptance for 200 GeV Pythia jets using a simplified CNN (see text) applied to various
image sizes. The deep CNN was applied to the $43\times43$ color images and the gluon jet efficiency at 50\% quark jet acceptance was 5.0\% (slightly worse than the 4.6\% for $33 \times 33$ images in Table~\ref{tab:effs}).}
\label{tab:gridsize}
\end{table}

\subsection{Image-size dependence}
Additional pixel sizes for the jet images were also considered. As the deep CNN architecture illustrated in Figure~\ref{fig:netarch} cannot be applied to significantly smaller input sizes due to the maxpooling layers, a simpler architecture was used: a single convolutional layer with 24 filters of size $6\times 6$, a $2\times2$ maxpooling, and an L2-regularization parameter of $10^{-6}$. The log transformation was included in the pre-processing for training this shallow network. The simplified networks were trained in the same way as the deep convolutional neural network. Table~\ref{tab:gridsize} shows the gluon jet efficiency at 50\% quark jet acceptance for this network trained on \SI{200}{GeV} Pythia jets discretized into $13\times13$, $23\times 23$, $33\times 33$, and $43\times 43$ grid sizes. The performance is robust over different pixelization schemes, with a decrease in performance for the smallest $13\times 13$ pixelization. 

We also tried the full deep convolutional neural network with $43\times 43$ color input images, finding $\varepsilon_g=5.0\%$ at $\varepsilon_q = 50\%$. This is slightly worse than the $\varepsilon_g= 4.6\%$ found for the same sample produced using $33\times33$ pixel inputs. 

\subsection{Event generator dependence}
Many studies have shown that the two event generators Pythia and Herwig produce significantly different quark and gluon jets~\cite{Gallicchio:2012ez,Badger:2016bpw,Aad:2014gea}. For example, the ATLAS collaboration performed a study on light-quark and gluon jet discrimination, comparing the performance of discriminants on Pythia- and Herwig-generated events to their performance on data. Quark jets and gluon jets were found to be easier to distinguish in Pythia and harder to distinguish in Herwig. The performance of the considered discriminants on data tended to be between their performance on Pythia and Herwig~\cite{Aad:2014gea}.

To explore the generator-dependence of our results, \SI{200}{GeV} and \SI{1000}{GeV} samples were generated with Pythia and Herwig. The two event generators were found to give similar quark distributions and disagree primarily on the gluon distributions. The baseline jet variables and the  convolutional network all indeed have worse performance on Herwig jets than on Pythia jets. A comparison of the discrimination power of the observables between the two generators is included in Table~\ref{tab:effs}.

\begin{figure}
\centering
\includegraphics[scale=0.8]{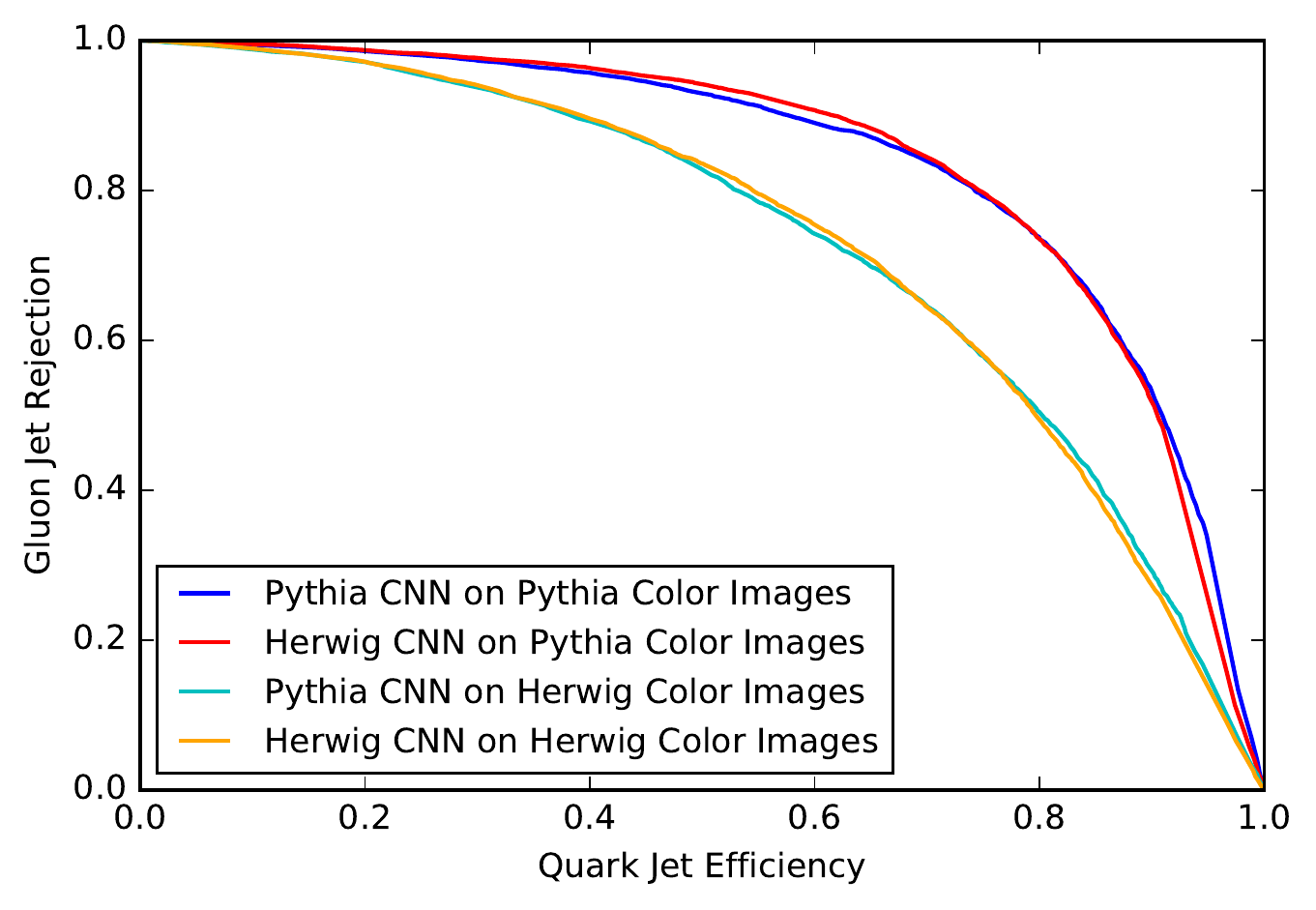}
\caption{ ROC curves for the Pythia- and Herwig-trained CNNs applied to  200~GeV samples generated with both of the generators.
Remarkably, the network performance seems robust to which samples are used for training.}
\label{fig:mc_roc}
\end{figure}

\begin{figure}
\centering
\includegraphics[scale=0.8]{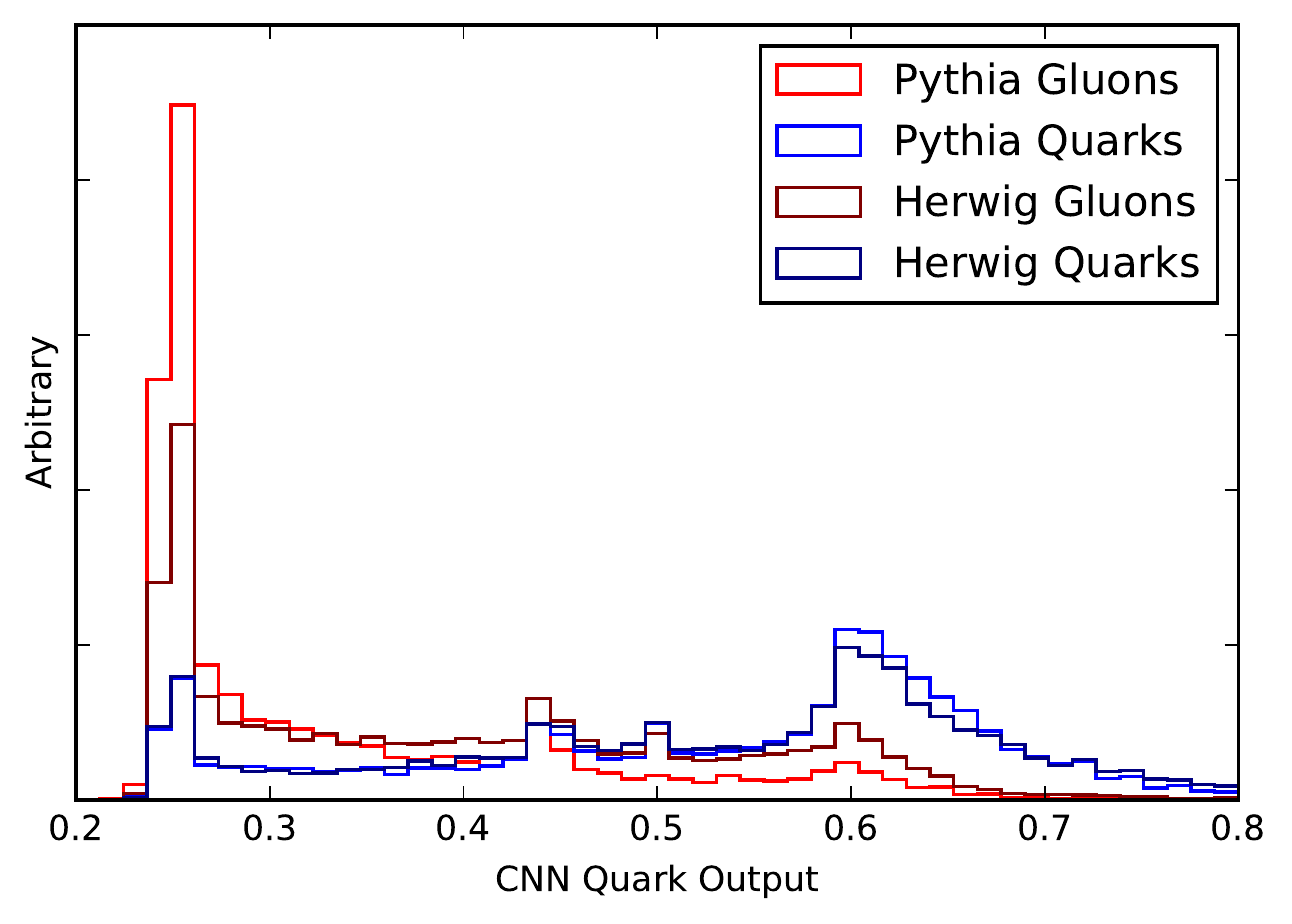}
\caption{Distribution of Pythia-trained colored-CNN outputs for 200 GeV jets. Quark samples are blue and gluon samples are red. Pythia samples are shown in the lighter shade and Herwig samples are shown in the darker shade. 
}
\label{fig:mc_hist}
\end{figure}

It is interesting to consider the four possibilities of applying the convolutional networks trained on Pythia jets or Herwig jets to test samples of Pythia jets or Herwig jets. Figure~\ref{fig:mc_roc} and Figure~\ref{fig:mc_hist} show the resulting ROC curves and distributions of convolutional network outputs on the colored jet images. We find that the network is surprisingly insensitive to the generator: the convolutional network trained on Pythia jets and tested on Herwig jets has comparable performance to the convolutional network trained directly on Herwig jets and tested on Herwig jets. This insensitivity is a positive sign for being able to train the network on MC-generated jets and apply it to data robustly.

\section{Conclusions \label{sec:conc}}
The ability to distinguish quark-initiated jets from gluon-initiated jets would be of tremendous practical application at colliders like the LHC. For example, many signals of beyond the standard model physics contain mostly quark jets, while their backgrounds are gluon-jet dominated. Quark/gluon jet discrimination is also extremely challenging: correlations in their radiation patterns and non-pertubative effects like hadronization are hard to disentangle. Thus this task is ideally suited for artificial intelligence. 

In this paper, we have applied machine learning techniques developed for computer vision, namely deep convolutional artifical neural networks, to the quark/gluon differentiation problem. Overall, we find excellent performance of the deep networks. In particular, these networks, which use essentially no input about the physics underlying the differences between these two jet types, performs as well or better than a collection of the best physically motivated observables from other studies (see Table~\ref{tab:effs}). 

The input layer of our neural network is taken to be the transverse momentum deposited in a particular region of the detector. We preserve the locality of this energy deposition by constructing a 2 dimensional jet image, with the 2 dimensions corresponding to the surface of a cylinder, in pseudorapidity and azimuthal angle (see Figure~\ref{fig:netarch}). It is not completely obvious that locality in position should be helpful in quark/gluon discrimination. On the one hand, locality is clearly relevant, as infrared safety demands that observables include integrations over compact phase space regions.  On the other hand, quarks and gluons are not infrared-safe objects; indeed hadronization, which depends on the color charge of the fragmenting partons, is non-local. The relevance of locality is essentially a consequence of the collinear singularity in the QCD splitting functions and of soft color coherence. Convolutional network architectures are structured to take advantage of the local information in the input while being able to learn non-local observables.

In addition to using the overall energy deposited in a local region, our network also exploits correlations in the particle charge and particle multiplicity. To use this information, we treat the input as a colored jet image. The red image is the transverse momenta in charged particles, the blue image is the transverse momenta of neutral particles, and the green image is the (local) charged particle multiplicity. Rather than colors, one can also think of these extra inputs as making our image 3 dimensional, with the third dimension only 3 pixels deep. The additional information gives a marginal improvement at 200 GeV, but for very high $p_T$ jets, the improvement is substantial. This is probably due to the overall higher multiplicity at higher energies and correspondingly enhanced discrimination power of observables that exploit this information (this can also be seen in the improved relative performance of charged-particle multiplicity at high $p_T$ in Table~\ref{tab:effs}).

A standard criticism of neural networks is that they are only as smart as the data used to train them is accurate: if the simulations are poor, it's garbage-in-garbage-out. For quark and gluon jet discrimination, this criticism must be taken seriously, as the simulations are known to be poor. Indeed, the two standard generators, Pythia and Herwig, differ substantially in their jet simulations with Herwig quark and gluon jets substantially more similar to each other than Pythia quark and gluon jets. We confirm this observation, as our NNs are subtantially worse at correctly labelling Herwig jets than Pythia jets. However, somewhat surprisingly, we find that the network performance is the same whether trained on a Pythia or Herwig sample. In other words, the networks may be picking up on underlying physical distinctions between the jets which are similar in the two simulations, just to a lesser degree. Essentially, the NN, when trained on either sample, is acting much like a physically-motivated observable. This suggests that the NN may be more trustworthy when acting on data than the simulations used to train them.

A second criticism of neural networks used in physics is that they are black boxes. They could be picking up on unphysical features of the simulations. While it is impossible to completely refute this criticism, it is certainly possible to explore what the networks are doing. For example, the convolutional network layers comprise filters which can be examined. An example of such filters is given in Fig.~\ref{fig:shallowDNNfilters} which shows how the NN is picking up on different elements of the radiation distribution. Also we can add merge layers, as described in Section~\ref{sec:merge}, which allow us to determine if a particular observable has been learned by the network. In fact, any multivariate technique, such as a boosted decision tree constructed from ten well-motivated observables, can be attacked with the black-box criticism. The truth is that because of the subtlety of the identification tasks we need these multivariate methods to perform, some lack of low-level understanding of what these methods are doing is unavoidable. Not understanding them does not mean the methods do not work. 

As particle colliders push to higher and higher luminosity, and as the signals we search for become ever more complex and subtle, the reliance on machine learning will eventually be inevitable. Indeed, machine learning has a bright future in particle physics, as developments in processing power, computer science and in event-generator simulations, give us more and more reason to use these methods. Moreover, artificial intelligence has the advantage of not being limited by human creativity. For example, a comprehensive quark/gluon discrimination study in~\cite{Gallicchio:2012ez} classified useful observables as being either ``shapes" or ``counts". However, another study~\cite{Pumplin:1991kc} found that $N_{90}$, a hybrid shape/count observable, works exceedingly well. One conclusion of our study is that deep neural networks can perform as well as any of the considered physical observables, or any combination thereof. Although we are not recommending that these networks be used on data yet (since the simulations on which they are trained cannot yet be trusted), our study does suggest that artificial intelligence will eventually play an essential role in quark/gluon jet discrimination and probably many other tasks in particle physics.   

\acknowledgments

The authors would like to thank Jesse Thaler, Christopher Frye, and Fr\'{e}d\'{e}ric Dreyer for useful discussions. PTK and EMM would like to thank the MIT Physics Department for its support. MDS was supported in part by the U.S. Department of Energy under grant DE-SC0013607. The computations in this paper were run on the Odyssey cluster supported by the FAS Division of Science, Research Computing Group at Harvard University.


\begin{thebibliography}{99}

\bibitem{Aad:2014yva} 
    ATLAS Collaboration, 
    \emph{A neural network clustering algorithm for the ATLAS silicon pixel detector}, 2014
    \emph{JINST} {\bf 9} P09009
    [\href{https://arxiv.org/abs/1406.7690}{{\tt arXiv:1406.7690}}].

\bibitem{Aad:2015ydr} 
    ATLAS Collaboration,
    \emph{Performance of $b$-Jet Identification in the ATLAS Experiment}, 2016
    \emph{JINST} {\bf 11} P04008
    [\href{https://arxiv.org/abs/1512.01094}{{\tt arXiv:1512.01094}}].

\bibitem{Tosi2015}
    M. Tosi, 
    \emph{Performance of Tracking, b-tagging and Jet/MET reconstruction at the CMS High Level Trigger},
    \emph{J. Phys. Conf. Ser.} {\bf 664} (2015).

\bibitem{CMS2012a}
    CMS collaboration,
    \emph{Performance of $\tau$-lepton reconstruction and identification in CMS}, 2012
    \emph{JINST} {\bf 7} P01001
    [\href{https://arxiv.org/abs/1109.6034}{\tt arXiv:1109.6034}].
    
\bibitem{Ball:2014uwa} 
    NNPDF Collaboration, R.D. Ball et al.,
    \emph{Parton distributions for the LHC Run II},
    \emph{JHEP} {\bf 04} (2015) 040
    [\href{https://arxiv.org/abs/1410.8849}{\tt arXiv:1410.8849}].
  
\bibitem{Gallicchio:2010dq} 
    J.~Gallicchio, J.~Huth, M.~Kagan, M.~D.~Schwartz, K.~Black and B.~Tweedie,
    \emph{Multivariate discrimination and the Higgs + W/Z search},
    \emph{JHEP} {\bf 04} (2011) 069
    [\href{https://arxiv.org/abs/1010.3698}{\tt arXiv:1010.3698}].

\bibitem{Maggipinto:1997eh} 
    T.~Maggipinto et al.,
    \emph{Role of neural networks in the search of the Higgs boson at LHC},
    \emph{Phys.\ Lett.} {\bf B 409} (1997) 517
    [\href{https://arxiv.org/abs/hep-ex/9705020}{\tt arXiv:9705020}].

\bibitem{CMS2013}
    CMS collaboration,
    \emph{Search for supersymmetry in events with opposite-sign dileptons and 
               missing transverse energy using an artificial neural network},
    \emph{Phys. Rev.} {\bf D 87} (2013) 072001
    [\href{https://arxiv.org/abs/1301.0916}{\tt arXiv:1301.0916}].

\bibitem{Baldi:2014pta} 
    P.~Baldi, P.~Sadowski and D.~Whiteson,
    \emph{Enhanced Higgs Boson to $\tau^+\tau^-$ Search with Deep Learning},
    \emph{Phys.\ Rev.\ Lett.} {\bf 114} (2015) 111801 
    [\href{https://arxiv.org/abs/1410.3469}{\tt arXiv:1410.3469}].

%\bibitem{Banfi:2006hf} 
% A.~Banfi, G.~P.~Salam and G.~Zanderighi,
%\emph{Infrared safe definition of jet flavor},
%\emph{Eur.\ Phys.\ J.\ C} {\bf 47} (2006) 113,
%[\href{https://arxiv.org/abs/hep-ph/0601139}{\tt arXiv:0601139}].
  
\bibitem{Cogan:2014oua} 
    J.~Cogan, M.~Kagan, E.~Strauss and A.~Schwarztman,
    \emph{Jet-Images: Computer Vision Inspired Techniques for Jet Tagging},
    \emph{JHEP} {\bf 02} (2015) 118
    [\href{https://arxiv.org/abs/1407.5675}{\tt arXiv:1407.5675 }].

\bibitem{Almeida:2015jua} 
    L.~G.~Almeida, M.~Backovi\'{c}, M.~Cliche, S.~J.~Lee and M.~Perelstein,
    \emph{Playing Tag with ANN: Boosted Top Identification with Pattern Recognition},
    \emph{JHEP} {\bf 1507} (2015) 086
    [\href{https://arxiv.org/abs/1501.05968}{\tt arXiv:1501.05968}].

\bibitem{deOliveira:2015xxd} 
    L.~de Oliveira, M.~Kagan, L.~Mackey, B.~Nachman and A.~Schwartzman,
    \emph{Jet-images $-$ deep learning edition},
    \emph{JHEP} {\bf 07} (2016) 069
    [\href{https://arxiv.org/abs/1511.05190}{\tt arXiv:1511.05190}].

\bibitem{Baldi:2016fql} 
    P.~Baldi, K.~Bauer, C.~Eng, P.~Sadowski and D.~Whiteson,
    \emph{Jet Substructure Classification in High-Energy Physics with Deep Neural Networks},
    \emph{Phys.\ Rev} {\bf D 93} (2016) 094034
    [\href{https://arxiv.org/abs/1603.09349}{\tt arXiv:1603.09349}].

\bibitem{Guest:2016iqz} 
    D.~Guest, J.~Collado, P.~Baldi, S.~C.~Hsu, G.~Urban and D.~Whiteson,
    \emph{Jet Flavor Classification in High-Energy Physics with Deep Neural Networks},
    \emph{Phys. Rev.} {\bf D 94} (2016) 112002
    [\href{https://arxiv.org/abs/1607.08633}{\tt arXiv:1607.08633}].
    
\bibitem{Barnard:2016}
    J. Barnard, E.N. l Dawe, M.J. Dolan, and N. Rajcic, 
    \emph{Parton Shower Uncertainties in Jet Substructure Analyses with Deep Neural Networks} 
    [\href{https://arxiv.org/abs/1609.00607}{\tt arXiv:1609.00607}].

\bibitem{kriz2012}
    A. Krizhevsky, I. Sutskever, and G.E. Hinton,
    \emph{Imagenet classification with deep convolutional neural networks}, in the proceedings of
    \emph{Neural Information Processing Systems} (NIPS 2012), December 3-8, Lake Tahoe, U.S.A (2012).

\bibitem{Pumplin:1991kc} 
    J.~Pumplin,
    \emph{How to tell quark jets from gluon jets},
    \emph{Phys.\ Rev.} {\bf D 44} (1991) 2025.

\bibitem{Lonnblad:1990}
    L. L\"{o}nnblad, C. Peterson, and T. R\"{o}gnvaldsson, 
    \emph{Finding gluon jets with a neural trigger},
    \emph{Phys. Rev. Lett.} {\bf 65} (1990) 1321.

\bibitem{Acton:1993}
    OPAL Collaboration,
    \emph{A Study of differences between quark and gluon jets using vertex tagging of quark jets},
    \emph{Z. Phys.} {\bf C 58} (1993) 387.

\bibitem{Alexander:1991}
    OPAL Collaboration,
    \emph{A direct observation of quark-gluon jet differences at LEP},
    \emph{Phys. Lett.} {\bf B 265} (1991) 462.

\bibitem{deLime:2016}
    D. Ferreira de Lima, P. Petrov, D. Soper, and M. Spannowsky,
    \emph{Quark-Gluon tagging with Shower Deconstruction: Unearthing dark matter and Higgs couplings}
    \href{https://arxiv.org/abs/1607.06031}{\tt arXiv:1607.06031}].

\bibitem{Larkoski:2014}
    A.J. Larkoski, J. Thaler, and W.J. Waalewijn, 
    \emph{Gaining (mutual) information about quark/gluon discrimination},
    \emph{JHEP} {\bf 11} (2014) 129 
    [\href{https://arxiv.org/abs/1408.3122}{\tt arXiv:1408.3122}].

\bibitem{Bhattacherjee:2015}
    B. Bhattacherjee, S. Mukhopadhyay, M.M. Nojiri, Y. Sakaki, and B.R. Webber,
    \emph{Associated jet and subjet rates in light-quark and gluon jet discrimination},
    \emph{JHEP} {\bf 04} (2013) 090 
    [\href{https://arxiv.org/abs/1501.04794}{\tt arXiv:1501.04794}].
    
\bibitem{Gallicchio:2012ez} 
     J.~Gallicchio and M.~D.~Schwartz,
    \emph{Quark and gluon jet substructure},
    \emph{JHEP} {\bf 04} (2013) 090
    [\href{https://arxiv.org/abs/1211.7038}{\tt arXiv:1211.7038}].

\bibitem{Gallicchio:2011xq} 
    J.~Gallicchio and M.~D.~Schwartz,
    \emph{Quark and gluon tagging at the LHC},
    \emph{Phys.\ Rev.\ Lett.} {\bf 107} (2011) 172001 
    [\href{https://arxiv.org/abs/1106.3076}{\tt arXiv:1106.3076}].

\bibitem{Badger:2016bpw} 
    J.R.~Andersen et al.,
    \emph{Les Houches 2015: Physics at TeV Colliders Standard Model Working Group Report} 
    [\href{https://arxiv.org/abs/1605.04692}{\tt arXiv:1605.04692}].

\bibitem{Aad:2014gea} 
    ATLAS Collaboration,
    \emph{Light-quark and gluon jet discrimination in $pp$ collisions at $\sqrt{s}=7\mathrm {\ TeV}$ 
               with the ATLAS detector},
    \emph{Eur.\ Phys.\ J.} {\bf C 74} (2014) 3023
    [\href{https://arxiv.org/abs/1405.6583}{\tt arXiv:1405.6583}].
    
\bibitem{Frye:2016aiz} 
    C.~Frye, A.J.~Larkoski, M.D.~Schwartz and K.~Yan,
    \emph{Factorization for groomed jet substructure beyond the next-to-leading logarithm},
    \emph{JHEP} {\bf 07} (2016) 064,
    [\href{https://arxiv.org/abs/1603.09338}{\tt arXiv:1603.09338}].

\bibitem{coll2008}
    R. Collobert and J. Weston,
    \emph{A unified architecture for natural language processing: deep neural networks with multitask learning},
    in the proceedings of the 25$^{\text{th}}$
    \emph{International Conference on Machine Learning}, July 5-9, Helsinki, Finland (2008).

\bibitem{bald2014}
    P. Baldi, P. Sadowski, and D. Whiteson,
    \emph{Searching for exotic particles in high-energy physics with deep learning},
    \emph{Nature Commun.} {\bf 5} (2014) 4308
    [\href{https://arxiv.org/abs/1402.4735}{\tt arXiv:1402.4735}].

\bibitem{diel2015}
    S. Dieleman, K.W. Willet, and J. Dambre,
    \emph{Rotation-invariant convolutional neural networks for galaxy morphology prediction}, 
    \emph{Mon. Roy. Astron. Soc.} {\bf 450} (2015) 1441
    [\href{https://arxiv.org/abs/1503.07077}{\tt arXiv:1503.07077}].

\bibitem{niel2015}
    M.A. Nielsen, 
    \emph{Neural Networks and Deep Learning}, 
    Determination Press, U.S.A (2015).

\bibitem{good2016}
I.~Goodfellow, Y.~Bengio, and A.~Courville, \emph{Deep Learning,} MIT Press, U.S.A. (2016).

\bibitem{relu}
X.~Glorot, A.~Bordes, and Y.~Bengio, \emph{Deep sparse rectifier neural networks}, proceedings of the 14$^{\text{th}}$ \emph{International Conference on Artificial Intelligence and Statistics}, April 11-13, Ft. Lauderdale, U.S.A. (2011).

\bibitem{horn1989}
K.~Hornik, M.~Stinchcombe, and H.~White, \emph{Multilayer feedforward networks are universal approximators, Neural Netw.} {\bf 2} (1989) 359.

\bibitem{schm2015}
J.~Schmidhuber, \emph{Deep learning in neural networks: an overview, Neural Netw.} {\bf 61} (2015) 85  [\href{https://arxiv.org/abs/1404.7828}{\tt arXiv:1404.7828}].

\bibitem{szeg2015}
C.~Szegedy et al., \emph{Going deeper with convolutions}, in the proceedings \emph{IEEE Conference on Computer Vision and Pattern Recognition}, June 7-12, Boston U.S.A. (2015) [\href{https://arxiv.org/abs/1409.4842}{\tt arXiv:1409.4842}].

\bibitem{dropout}
N. Srivastava et al., \emph{Dropout: a simple way to prevent neural networks from overfitting, JMLR} {\bf 15} (2014) 1929.

 \bibitem{Pythia8.2:2015}
 T. Sjostrand, et al., \emph{An Introduction to PYTHIA 8.2, Comput. Phys. Commun.} {\bf 191} (2015) 159, [\href{https://arxiv.org/abs/1410.3012}{\tt arXiv:1410.3012}].
 
\bibitem{Herwig:2008}
 M. Bahr et al.,, \emph{Herwig++ physics and manual, Eur. Phys. J.} {\bf C 58} (2008) 639 [\href{https://arxiv.org/abs/0803.0883}{\tt arXiv:0803.0883}].
 
 \bibitem{Herwig:2016}  
 J.~Bellm et al., \emph{Herwig 7.0 / Herwig++ 3.0 release note, Eur. Phys. J.} {\bf C 76} (2016) 196  [\href{https://arxiv.org/abs/1512.01178}{\tt arXiv:1512.01178}].
 
 \bibitem{FastJet:2012}
 M.~Cacciari, G.P.~Salam, and G.~Soyez, \emph{FastJet user manual, Eur. Phys. J.} {\bf C 72} (2012) 1896 [\href{https://arxiv.org/abs/1111.6097}{\tt arXiv:1111.6097}].

\bibitem{cacc2008}
M.~Cacciari, G.P.~Salam, and G.~Soyez, \emph{The anti-$k_t$ jet clustering algorithm, JHEP} {\bf 04} (2008) 063  [\href{https://arxiv.org/abs/0802.1189}{\tt arXiv:0802.1189}].

\bibitem{Kaplan:2008ie} 
  D.~E.~Kaplan, K.~Rehermann, M.~D.~Schwartz and B.~Tweedie, \emph{Top tagging: a method for identifying boosted hadronically decaying top quarks,
  Phys. Rev. Lett.} {\bf 101} (2008) 142001 
 [\href{https://arxiv.org/abs/0806.0848}{\tt arXiv:0806.0848}].
 
\bibitem{NNPDF:2009}
    NNPDF Collaboration, R.D. Ball et al.,
    \emph{A determination of parton distributions with faithful uncertainty estimation}, 
    \emph{Nucl. Phys.} {\bf B 809} (2009) 1 [\emph{Erratum ibid.} {\bf B 816} (2009) 293] 
    [\href{https://arxiv.org/abs/0808.1231}{\tt arXiv:0808.1231}].
    
\bibitem{simard:2003}
P.Y.~Simard, D.~Steinkraus, and J.C.~Platt, \emph{Best practices for convolutional neural networks applied to visual document analysis, ICDAR} {\bf 3} (2003).

\bibitem{keras}
F.~Chollet, \emph{Keras,} available at \href{https://github.com/fchollet/keras}{GitHub} (2015).

\bibitem{theano}
J.~Bergstra et al., \emph{Theano: a CPU and GPU math compiler in python}, in the proceedings of the \emph{$9^{\text{th}}$ Python in Science Conference}, June 28-July 3, Austin, U.S.A. (2010).

\bibitem{heuniform}
K.~He, X.~Zhang, S.~Ren, and J.~Sun, \emph{Delving deep into rectifiers: surpassing human-level performance on imagenet classification}, in the proceedings of the \emph{IEEE International Conference on Computer Vision}, December 11-18, Santiago, Chile (2015).

\bibitem{adam}
D.~Kingma and J.~Ba, \emph{Adam: A method for stochastic optimization,} \href{https://arxiv.org/abs/1412.6980}{\tt arXiv:1412.6980}.

\bibitem{spearmint}
J.~Snoek, H.~Larochelle, and R.P.~Adams, \emph{Practical bayesian optimization of machine learning algorithms}, in the proceedings of \emph{Neural Information Processing Systems (NIPS 2012)}, December 3-8, Lake Tahoe, U.S.A. (2012), \href{https://arxiv.org/abs/1206.2944}{\tt arXiv:1206.2944}.

\bibitem{CMS:2009nxa} 
CMS Collaboration, \emph{Particle-flow event reconstruction in CMS and performance for jets, taus, and MET,} CMS-PAS-PFT-09-001 (2009).

\bibitem{sklearn}
F.~Pedregosa et al., \emph{Scikit-learn: machine learning in python, JMLR} {\bf 12} (2012) 2825.

\bibitem{Larkoski:2013}
A.J.~Larkoski, G.P~Salam, and J.~Thaler, \emph{Energy correlation functions for jet substructure}, \emph{JHEP} {\bf 06} (2013) 108 [\href{https://arxiv.org/abs/1305.0007}{\tt arXiv:1305.0007}].

\bibitem{Kucuk:2016}
H.~K\"{u}\c{c}\"{u}k, \emph{Measurement of the inclusive-jet cross-section in proton-proton collisions and study of Quark-Gluon Jet discrimination with the ATLAS experiment at the LHC}, Dissertation, University College London, London. U.K. (2016).

\end{thebibliography}
\end{document}